\RequirePackage{lineno}
\documentclass[aps,showpacs,preprintnumbers,amsmath,nofootinbib,amssymb]{revtex4}
\usepackage[utf8]{inputenc}
\usepackage{graphicx}
\usepackage{mathrsfs}
\usepackage{bm}
\usepackage{hyperref}
\usepackage{indentfirst}
\usepackage{epstopdf}
\usepackage{color}

\newcommand{\dd}{\mathrm{d}}
\usepackage{blindtext}
\begin{document}
 %\linenumbers

\title{Stochastic reconstructions of spectral functions: Application to lattice QCD}

\author{
   H.-T. Ding$^{\rm a}$, O. Kaczmarek$^{\rm a,b}$,
	Swagato Mukherjee$^{\rm c}$, H. Ohno$^{\rm c,d}$, H.-T. Shu$^{\rm a}$
}

%\address{
\affiliation{
	$^{\rm a}$Key Laboratory of Quark and Lepton Physics (MOE) and Institute of
	Particle Physics, \\
	Central China Normal University, Wuhan 430079, China\\
	$^{\rm b}$Fakult\"at f\"ur Physik, Universit\"at Bielefeld, D-33615 Bielefeld,
	Germany \\
	$^{\rm c}$Physics Department, Brookhaven National Laboratory, 
	Upton, New York 11973, USA\\
		$^{\rm d}$Center for Computational Sciences, University of Tsukuba, Tsukuba, Ibaraki 305-8577, Japan 
}

\begin{abstract}
 We present a detailed study of the applications of two stochastic approaches, stochastic optimization method (SOM) and stochastic analytical inference (SAI), to extract spectral functions from Euclidean correlation functions. SOM has the advantage that it does not require prior information. On the other hand, SAI is a more generalized method based on Bayesian inference. Under mean field approximation SAI reduces to the often-used maximum entropy method (MEM), and for a specific choice of the prior SAI becomes equivalent to SOM. To test the applicability of these two stochastic methods to lattice QCD, firstly, we apply these methods to various reasonably chosen model correlation functions, and present detailed comparisons of the reconstructed spectral functions obtained from SOM, SAI and MEM. Next, we present similar studies for charmonia correlation functions obtained from lattice QCD computations using clover-improved Wilson fermions on large, fine, isotropic lattices at $0.75$ and $1.5T_c$, $T_c$ being the deconfinement transition temperature of a pure gluon plasma. We find that SAI and SOM give consistent results to MEM at these two temperatures. 
 
% However, by analyzing the reconstructed correlators we find that stochastic methods have larger dependences on the temporal extent at the current noise level.
% We find that SAI and SOM give consistent results that suggest dissociation of $\eta_c$ and $J/\psi$ in the gluon plasma already at $1.5T_c$. These findings reinforce the previous conclusions of Ref.~\cite{Ding:2012sp} that was solely based on MEM.

\end{abstract}
\pacs{12.38.Gc, 12.38.Mh, 25.75.Nq, 25.75.-q}

\maketitle
\section{Introduction}
\label{introduction}

One of the central goals of the physics program of the present and the future
heavy ion colliders is the exploration of the phase diagram and transport properties of strongly
interacting matter. At vanishing baryon chemical potential the QCD transition from hadronic phase to QGP phase is predicted not to be a
real phase transition but an analytic rapid crossover~\cite{Aoki:2006we,Bazavov:2011nk}, and its chiral and deconfinements aspects can be reflected by thermal modifications of light and heavy hadrons, while the transport properties of the medium are related to the propagation of conserved currents.
The dilepton spectrum covering the mass region of $\rho$ meson~\cite{Brandenburg:2017meb,Adamczyk:2015lme,Adare:2015ila} and the suppression of the yields of heavy quarkonia as well as open charm/bottom hadrons in the $\rm PbPb/AuAu$ collisions compared to those in the $\rm pp$ collisions have been extensively studied in the experiments at RHIC and LHC~\cite{ALICE:2012ab,Adare:2006ns,Chatrchyan:2012np}. Connecting these experimental observations to fundamental interactions of QCD requires a thorough understanding of the in-medium modifications of hadrons and transport properties such as heavy quark diffusion coefficients. Theoretically the key point is the hadron spectral function as which encodes all the information about the hadron. Besides that the spectral function in the vector channel is related to the thermal dilepton production rate \cite{Braaten:1990wp} and its low frequency part also gives transport coefficients such as the electrical conductivity and the heavy quark diffusion coefficient through Kubo formulas. Thus by investigating on the change of resonance peak structure and the slope at the vanishing frequency in the spectral function at various temperatures it is possible to determine the dissociation temperature of hadrons and the diffusion coefficients, respectively.

First principle lattice QCD has been a useful tool to study  
the in-medium properties of hadrons as well as the transport properties of the medium~\cite{Ding:2015ona}. However, despite the importance of the spectral functions to understand in-medium behaviors of the strong interaction matters, the spectral functions cannot be calculated directly using lattice QCD. Instead, what one can calculate is the Euclidean correlation function, $G$, which is related to the spectral function, $\rho$, as
\begin{equation}
\label{correlator_spf}
G(\tau,T)= \int_0^{\infty} \frac{d \omega}{2\pi}\rho(\omega,T) K(\omega,\tau,T),
\end{equation}
where the integration kernel at finite temperature, $K(\omega,\tau,T)$, is given, e.g. in the bosonic case as
\begin{equation}
\label{kernel}
K(\omega,\tau,T) \equiv \frac{\cosh(\omega( \tau-\frac{1}{2T}))}{\sinh(\frac{\omega}{2T})}.
\end{equation}
To extract the spectral function from the correlation function one needs to solve an ill-posed inverse problem. Practically, the correlation function is only given at $\mathcal{O}(10)$ discrete imaginary-time distances, $\tau$, with some errors while, at least $\mathcal{O}(1000)$ data points in frequency, $\omega$, are needed for sufficiently good resolution of the spectral function. This is a typical ill-posed problem, where degrees of freedom of the input are much smaller than the output, leading to infinite number of possible solutions. Therefore, in general, a simple {$\chi^2$}-fitting is not applicable unless sufficiently detailed prior information on the spectral function is known \cite{Ghiglieri:2016tvj,Ding:2016hua}. This is a reason why various methods have been developed to tackle this problem. A most commonly used method to date is the maximum entropy method, where the most likely solution based on the Bayes’ theorem can be selected, and has been introduced to lattice QCD studies~\cite{Asakawa:2000tr,Aarts:2007pk,Ikeda:2016czj}. Recently, a new Bayesian approach similar to MEM by replacing the Shannon-Jaynes entropy with a different term has been also proposed~\cite{Burnier:2013nla}. Some other methods like the Backus Gilbert method \cite{Francis:2015daa, Brandt:2015aqk}, which manipulates in the local vicinity of some frequency range in a model-independent way, and the Tikhonov method with Morozov discrepancy principle \cite{Dudal:2013yva}, have been presented as well.

An advantage of the Bayesian methods like MEM is that they guarantee a unique solution under certain prior information, which allows us to overcome ill-posed problems. However, this leads to uncertainties depending on the prior information. Therefore, one should check the uncertainties carefully by changing prior information and, by comparing results between as many different methods as possible. In this paper we make use of two stochastic approaches to extract spectral functions in lattice QCD calculations, namely the stochastic analytical inference (SAI) \cite{PhysRevE.81.056701, 2004cond.mat..3055B} and the stochastic optimization method (SOM) \cite{PhysRevB.62.6317}. The key idea behind these methods is to use Monte Carlo averages over a wide range of possible spectra weighted by a certain criteria instead of selecting the most probable solution as for the MEM. Our goal is to examine the suitability of these stochastic methods for lattice QCD and provide a more robust estimate of the the systematic uncertainties of the spectral functions obtained from lattice QCD calculations. For this purpose, we focus on the reanalyses of the charmonia spectral functions in gluon plasma previously presented in Ref.~\cite{Ding:2012sp}.

The rest of the paper is organized as follows. In Sec.~\ref{basics_of_stochastic_approaches} we introduce the stochastic approaches and clarify the relationships among SAI, MEM and SOM. In Sec.~\ref{Implemention} detailed numerical implementations of the stochastic methods are given. In Sec.~\ref{mock_data_test} we test the methods with various model data which mimic possible charmonium spectral functions expected at several different cases. Then, we apply these methods to extract spectral functions from charmonium correlation functions computed using lattice QCD simulations at a finite temperature in Ref.~\cite{Ding:2012sp} in Sec.~\ref{real_data_analysis}. Finally, we summarize our results in Sec.~\ref{conclusion}.

\section{Basics of Stochastic Approaches}
\label{basics_of_stochastic_approaches}
In stochastic approaches a sequence of possible spectra are generated stochastically, and their average is taken. SAI \cite{PhysRevE.81.056701, 2004cond.mat..3055B} gives an averaged spectral image weighted by probability derived using Bayesian inference similar to MEM, which depends on prior information of the spectral function. On the other hand, SOM \cite{PhysRevB.62.6317} also takes an average over all possible spectra but without any prior knowledge as inputs. In the following sections we review the basics of stochastic approaches and also show relations among SAI, SOM and MEM.

\subsection{Bayesian statistical inference}
\label{Bayesian_statistical_inference}
First we start from the Bayesian statistical inference embedded in SAI following Ref.~\cite{PhysRevE.81.056701}. Suppose we try to extract a spectral image, $\rho$, from correlation function data, $G$, with a given prior knowledge or so-called default model (DM), $D$, where $D$ contains some information about the spectral function such as positivity. Here we also introduce a regularization parameter, $\alpha$, which controls contributions to the reconstructed image from the prior information relative to the data. According to the Bayes' theorem, $P[\rho|G,D,\alpha]$, the conditional probability having $\rho$ with given $G$, $D$ and $\alpha$, can be written by
\begin{equation}
\label{posterior_probability}
P[\rho|G,D,\alpha] = \frac{P[G|\rho,D,\alpha]P[\rho|D,\alpha]}{P[G|D,\alpha]},
\end{equation}
where $P[G|\rho,D,\alpha]$ and $P[\rho|D,\alpha]$ are the likelihood function and the prior probability, respectively. $P[G|D,\alpha]$ is a $\rho$-independent normalization. Once $P[\rho|G,D,\alpha]$ is calculated, the average over all possible spectra weighted by $P[\rho|G,D,\alpha]$ is given as
\begin{equation}
\label{average_spf}
\langle \rho \rangle_\alpha = \int \mathcal{D}\rho\;\rho\;P[\rho|G,D,\alpha].
\end{equation}
Then, a final image is given after eliminating the dependence on $\alpha$ by taking another weighted average over $\alpha$ as
\begin{equation}
\label{final_spf_SAI}
\left\langle\langle \rho \rangle\right\rangle=\int \dd \alpha\;\langle \rho \rangle_\alpha\;P[\alpha|G, D],
\end{equation}
where using the Bayes' theorem again, the conditional probability $P[\alpha|G, D]$ can be written by
\begin{equation}
\label{posterior_alpha}
P[\alpha|G, D] = \frac{P[G|D, \alpha]P[\alpha|D]}{P[G|D]} = \frac{P[\alpha|D]}{P[G|D]} \int \mathcal{D}\rho\;P[G|\rho,D,\alpha]P[\rho|D,\alpha].
\end{equation}

One can also study statistical uncertainties of the reconstructed image. Since there are correlations among $\rho(\omega)$ at different frequencies, following  Refs.~\cite{Asakawa:2000tr,Jarrell:1996rrw} we introduce the spectral function averaged over a certain frequency range, $I$, as
\begin{equation}
\label{mean_alpha}
\begin{split}
\langle \bar{\rho}_I\rangle_{\alpha} &\equiv \frac{\int \mathcal{D}\rho \int_I \dd \omega\  \rho(\omega)P[\rho|G,D,\alpha]}{\int_I\dd \omega}\\
&=\frac{\langle \int_I \dd \omega\ \rho(\omega)\rangle_{\alpha}}{\int_I \dd \omega}\\
&=\frac{\int_I \dd \omega\ \langle \rho(\omega)\rangle_{\alpha}}{\int_I\dd \omega}.
\end{split}
\end{equation}
Then, the variance is given as
\begin{equation}
\label{variance_alpha}
\begin{split}
\langle (\delta\bar{\rho}_I)^2\rangle_{\alpha} &\equiv \frac{\int \mathcal{D}\rho \int_{I\times I} \dd \omega \dd \omega’\ \delta\rho(\omega)\delta\rho(\omega’)P[\rho|G,D,\alpha]}{\int_{I\times I}\dd \omega\dd \omega’}\\
&=\frac{\langle \int_{I\times I} \dd \omega \dd \omega’\ \delta\rho(\omega)\delta\rho(\omega’)\rangle_{\alpha}}{\int_{I\times I} \dd \omega\dd \omega’}\\
&=\frac{\int_{I\times I} \dd \omega\dd \omega’\ \langle \delta\rho(\omega)\delta\rho(\omega’)\rangle_{\alpha}}{\int_{I\times I}\dd \omega\dd \omega’},
\end{split}
\end{equation}
where $\delta\rho(\omega)\equiv \rho(\omega)-\langle \rho(\omega)\rangle_{\alpha}$. Finally, $\alpha$ dependence is eliminated as
\begin{equation}
\label{variance_average}
\begin{split}
&\big{\langle} \langle \bar{\rho}_I\rangle \big{\rangle}=\int \dd \alpha\ \langle \bar{\rho}_I\rangle_{\alpha}P[\alpha|G,D],\\
&\big{\langle} \langle (\delta\bar{\rho}_I)^2\rangle \big{\rangle}=\int \dd \alpha\  \langle (\delta \bar{\rho}_I)^2\rangle_{\alpha}P[\alpha|G,D].
\end{split}
\end{equation}
The above equations can be used to estimate the uncertainties in MEM as well as the stochastic methods to be discussed in the following sections. In MEM the probability $P[\rho|G,D,\alpha]$ is assumed to be a sharp Gaussian distribution. Thus the variance at a certain $\alpha$ can be approximated as~\cite{Asakawa:2000tr,Jarrell:1996rrw}
\begin{equation}
\label{variance_alpha_mem}
\langle (\delta\bar{\rho}_I)^2\rangle_{\alpha}^{MEM} \approx -\int_{I\times I} \dd \omega\dd \omega’\ \Big{(}\frac{\delta^2 Q}{\delta \rho(\omega) \delta \rho(\omega’)}\Big{)}_{\rho=\rho_{\alpha}}^{-1}\Big{/}\int_{I\times I}\dd \omega\dd \omega’,
\end{equation}
where the definition of Q can be found in Sec.\ref{SAI_to_MEM}.

\subsection{Stochastic analytical inference}
\label{Stochastic_Analytical_Inference}
Following Ref.~\cite{2004cond.mat..3055B}, in this section we show how to specify the explicit forms of the probabilities mentioned above in SAI. First, for convenience, let us introduce the modified spectral function, $\tilde \rho(\omega)\equiv \rho(\omega)K(\omega,\tau_0)$, the modified DM, $\tilde D(\omega)=D(\omega)K(\omega,\tau_0)$ and the modified kernel $\tilde K(\omega,\tau)\equiv K(\omega,\tau)/K(\omega,\tau_0)$\footnote{In general, this is not necessary but it can avoid divergence in the kernel Eq.(\ref{kernel}) at $\omega=0$ and also allows to have a simple normalization condition Eq.(\ref{normalization_of_field}) without $\tilde K$ dependence.}, where $\tau_0$ is a reference imaginary time. As Beach proposed in Ref.~\cite{2004cond.mat..3055B}, a mapping from frequency, $\omega$, onto a new variable, $x\in[0,x_\mathrm{max}]$
\begin{equation}
\label{mapping}
x\equiv\phi(\omega)=\int_0^{\omega}\ \frac{\dd \omega'}{2\pi}\;\tilde D(\omega')
\end{equation}
is introduced, where $\tilde D$ is positive definite and $x_\mathrm{max} \equiv \phi(\infty)$. By changing $\omega$ to $x$ in Eq.(\ref{correlator_spf}), the correlation function reconstructed from a given spectral function can be written as
\begin{equation}
\label{rewrite}
G_\mathrm{rec}(\tau)=\int_{0}^{x_{max}}\dd x\;n(x)\tilde{K}(\phi^{-1}(x),\tau),
\end{equation}
where
\begin{equation}
\label{field}
n(x) \equiv \frac{\tilde\rho(\phi^{-1}(x))}{\tilde D(\phi^{-1}(x))} = \frac{\rho(\phi^{-1}(x))}{D(\phi^{-1}(x))}.
\end{equation}
Consequently, the newly defined function $n(x)$ is normalized as
\begin{equation}
\label{normalization_of_field}
\int_{0}^{x_{max}}\dd x\;n(x)=G(\tau_0).
\end{equation}
Since $\rho$ can be calculated from the relation Eq.(\ref{field}) once $n(x)$ is given, from here on we consider reconstruction of $n(x)$ instead of $\rho(\omega)$ itself.

Suppose we have $N_\mathrm{conf}$ sets of correlator data, $\{G^i(\tau)\;|\;i=1,2,\cdots,N_\mathrm{conf}\}$, at $N$ data points, $\hat\tau=\hat\tau_\mathrm{min}, \hat\tau_\mathrm{min}+1, \cdots, \hat\tau_\mathrm{max}=\hat\tau_\mathrm{min}+N-1$, where $\hat\tau \equiv \tau/a$ with lattice spacing $a$. Here the mean value, $\overline G$, and the covariance matrix, $C$, are given by
\begin{equation}
\overline G(\tau) \equiv \frac{1}{N_\mathrm{conf}}\sum_{i=1}^{N_\mathrm{conf}}G^{i}(\tau),
\end{equation}
\begin{equation}
\label{covariance_matrix}
C(\tau,\tau') \equiv \sum_{i=1}^{N_\mathrm{conf}}\frac{(\overline G(\tau)-G^i(\tau))(\overline G(\tau')-G^i(\tau'))}{N_\mathrm{conf}\cdot(N_\mathrm{conf}-1)}.
\end{equation}
According to the central limit theorem the correlators are expected to be Gaussian distributed for sufficiently large $N_\mathrm{conf}$. Therefore, it is natural to have the likelihood function, $P[G|n,D,\alpha]$ ,as
\begin{equation}
\label{likelihood}
P[G|n, D, \alpha] = \frac{1}{Z}e^{-\chi^2[n]/\alpha},
\end{equation}
where
\begin{equation}
\label{target_function}
\chi^2=\frac{1}{2}\sum_{\hat\tau,\hat\tau'=\hat\tau_\mathrm{min}}^{\hat\tau_\mathrm{max}}(G_\mathrm{rec}(\tau)-\overline{G}(\tau))C^{-1}(\tau,\tau')(G_\mathrm{rec}(\tau')-\overline{G}(\tau')).
\end{equation}
The normalization factor, $Z$, can be computed as
\begin{equation}
\label{Z_alpha}
Z = \int \mathcal{D}\overline G\;e^{-\chi^2/\alpha} = (2\pi\alpha)^{N/2} \sqrt{\det C}.
\end{equation}
On the other hand, since we have the prior information of the spectral function through the mapping Eq.(\ref{mapping}), which is encoded into the normalization condition Eq.(\ref{normalization_of_field}),
the prior probability, $P[n|D,\alpha]$, should be given as
\begin{equation}
\label{prior_probability}
P[n|D,\alpha] = \delta \left(\int_0^{x_\mathrm{max}} \dd x\;n(x)-\overline G(\tau_0)\right).
\end{equation}
Then, the $n$-independent normalizaiton, $P[G|D, \alpha]$ can be written as
\begin{equation}
\label{evidence}
P[G|D,\alpha] = \frac{Z'}{Z} = \frac{1}{Z} \int\mathcal{D}'n\;e^{-\chi^2[n]/\alpha},
\end{equation}
where $\mathcal{D}'n \equiv \left(\prod_x \dd n(x) \right)\delta \left(\int_0^{x_\mathrm{max}} \dd x\;n(x)-\overline G(\tau_0)\right)$. As a result, replacing $\rho$ with $n$ in Eq.(\ref{average_spf}) and substituting Eq.(\ref{likelihood}), Eq.(\ref{prior_probability}) and Eq.(\ref{evidence}) into Eq.(\ref{posterior_probability}), one gets the following expression:
\begin{equation}
\label{average_SPF_SAI}
\langle n(x) \rangle_\alpha = \frac{1}{Z'}\int\mathcal{D}'n\;n(x)\;e^{-\chi^2[n]/\alpha}.
\end{equation}
Similarly, from Eq.(\ref{posterior_alpha}),
\begin{equation}
\label{posterior_alpha_SAI}
P[\alpha|G,D] \propto P[\alpha|D]\;\alpha^{-N/2}\int\mathcal{D}'n\;e^{-\chi^2[n]/\alpha},
\end{equation} 
where $\alpha^{-N/2}$ comes from Eq.(\ref{Z_alpha}). The probability $P[\alpha|D]$ is unknown. Conventionally, $P[\alpha|D]\propto$ 1 or $1/\alpha$ are chosen \cite{Jeffreys:1998,Box:1992}. However, the choice is irrelevant for the final results if the data size $N$ is sufficiently large (see Appendix.\ref{AppendixA}). To get Eq.(\ref{posterior_alpha_SAI}) explicitly one needs to calculate the partition function $Z'=\int\mathcal{D}'n\;e^{-\chi^2/\alpha}$. By introducing the density of states (DoS), $\Omega(E)=\int \mathcal{D}'n\;\delta(\chi^2[n]-E)$, $Z'$ can be rewritten as
\begin{equation}
\label{Z_change}
Z'=\int\dd E\;\Omega(E)\;e^{-E/\alpha}.
\end{equation}
Calculating $Z'$ thus is equivalent to calculating $\Omega(E)$. The DoS can be evaluated numerically by using, e.g. the Wang-Landau algorithm \cite{PhysRevE.90.023302}. The Wang-Landau algorithm is briefly reviewed in Appendix. \ref{AppendixB}. The final spectral function, $\left\langle\langle n(x) \rangle\right\rangle$, is given by taking an average of $\langle n(x) \rangle_\alpha$ weighted by Eq.(\ref{posterior_alpha_SAI}) over all $\alpha$
\begin{equation}
\label{final_spf_sai}
\left\langle\langle n(x) \rangle\right\rangle=\int \dd \alpha\ \langle n(x) \rangle_\alpha\ P[\alpha|G,D]\ .
\end{equation}

\subsection{Solution of SOM through the kink condition}
\label{SOM_condition}
SOM is another stochastic approach tackling the inversion problem. Different from SAI, SOM does not need any prior information about the solution. Thus there is no default model used and we do not introduce the coordinate mapping. The main idea of SOM is to average over all the independent possible solutions obtained using a modified simulated annealing algorithm (SAA). Similar to SAI, there are two quantities controlling the system, the fictitious temperature $\alpha$ which decreases exponentially to a quite small value $\alpha_{stop}$ and the internal energy $\chi^2[\rho]$. For a well-defined system, as the temperature $\alpha$ of the system decreases, the internal energy $\chi^2$ would definitely decrease in the same pattern and the optimal possible solution would appear when temperature goes to 0 if the system is detailed-balanced at each temperature. However, since our system is ill-posed, the simulation would be overfitted when $\alpha$ approaches 0. One way out is to sample the spectral functions before overfitting. And we call the point that the system starts to overfit a $kink$.

\begin{figure}[htb]
\centerline{\includegraphics[width=0.4\textwidth]{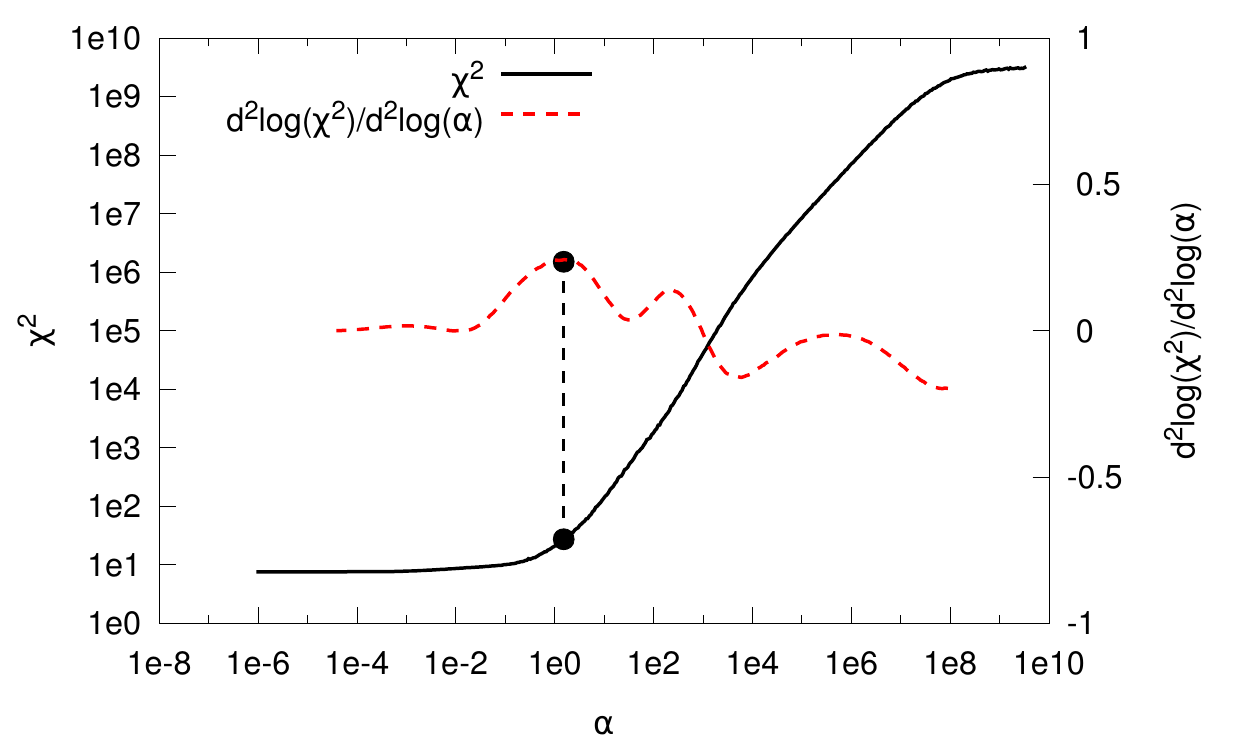} }
\caption{A typical structure of $\chi^2-\alpha$ curve and its $kink$ point.}
\label{kink}
\end{figure}

To find such a ``kink" point one can calculate $\log (\chi^2)$'s second derivative respect to $\log (\alpha)$ (taking the logarithm here is for convenience because $\alpha$ is decreased exponentially)  using quartic-basis spline fits. And the $kink$ is located at the maximum of the second derivative. We find that unless data points are very limited, we can always specify such a $kink$ point at some certain temperature $\alpha^*$ which picks out a unique solution. In Fig.\ref{kink} we show the kink obtained in one of our model data tests.  We can see that as $\alpha$ decreases from $10^{9}$ to $10^{-6}$, $\chi^2$ decreases following $\alpha$ in the range [$10^9$, $10$] but after that $\chi^2$ does not change much. And around the transition region $\alpha \sim 1$, a clear maximum appears in the $d^2\log(\chi^2)/d^2\log(\alpha)-\alpha$ curve which specifies the $kink$ point.  

\subsection{Relation between SAI and MEM and SOM}
\label{SAI_to_MEM}
\subsubsection{SAI to MEM}
The formalism of MEM can be derived by repeating the similar argument as in the previous section. Following \cite{Jarrell:1996rrw}, the likelihood function in MEM is given as
\begin{equation}
P[G|\rho,D,\alpha] = \frac{1}{Z_L}e^{-\chi^2[\rho]},
\end{equation}
where $Z_L$ is a normalization factor. On the other hand, for a given $\alpha$ the prior probability is explicitly obtained by the Shannon-Jaynes entropy
\begin{equation}
\label{entropy}
S[\rho] = -\int_0^\infty\frac{\dd\omega}{2\pi}\;\rho(\omega)\;\ln \frac{\rho(\omega)}{D(\omega)},
\end{equation}
as
\begin{equation}
P[\rho|D,\alpha] = \frac{1}{Z_S(\alpha)}e^{\alpha S[\rho]},
\end{equation}
where $Z_s(\alpha) \simeq (2\pi/\alpha)^{N/2}$. Therefore, the conditional probability, $P[\rho|G,D,\alpha]$, can be written as
\begin{equation}
P[\rho|G,D,\alpha] \propto \frac{1}{Z_LZ_S(\alpha)}e^{-Q_\alpha[\rho]},
\end{equation}
where $Q_\alpha\equiv \chi^2-\alpha S$. In contrast to SAI, where the average spectral function is calculated as Eq.(\ref{average_SPF_SAI}), MEM picks up only the most probable solution, $\rho_\alpha^*$, which maximizes $P[\rho|G,D,\alpha]$, or in other words, minimizes Q assuming $P[\rho|G,D,\alpha]$ is sharply peaked around $\rho_\alpha^*$. This means that $\rho_\alpha^*$ is given by solving the following equation:
\begin{equation}
\label{direvative_of_q}
\left.\frac{\delta Q[\rho]}{\delta \rho(\omega)}\right|_{\rho=\rho^*_{\alpha}}=0.
\end{equation}
From Eq.(\ref{posterior_alpha}) the final spectral function is expressed as
\begin{equation}
\label{final_SPF_MEM}
\begin{split}
\left\langle \langle \rho \rangle \right\rangle &\propto \int \dd \alpha P[\alpha|G,D] \int \mathcal{D}\rho\;\rho(\omega)\;e^{-Q_\alpha[\rho]}\\
&\simeq \int \dd \alpha\ \rho^*_{\alpha}(\omega)\ P[\alpha|G,D],
\end{split}
\end{equation}
where in the second step $P[\rho|G,D,\alpha]$ is again assumed to be sharply peaked around $\rho_{\alpha}^*(\omega)$ and
\begin{equation}
P[\alpha|G,D] \propto P[\alpha|D]\int \mathcal{D}\rho\frac{1}{Z_LZ_S(\alpha)}e^{-Q_\alpha[\rho]}
\end{equation}
can be also evaluated under this assumption.

A question here is what is the relation between the output spectra from SAI and MEM. Actually, it has been proved that SAI is a generalization of MEM and is formally equivalent to MEM in a certain limit\cite{2004cond.mat..3055B}. To see this let us focus on the spectral functions at a given $\alpha$ from SAI and MEM. As already seen above, the most probable image in MEM is given by a solution of Eq.(\ref{direvative_of_q}) or equivalently a self-consistent equation as follows:
\begin{equation}
\label{mem_image}
\tilde \rho^*_{\alpha}(\omega)=e^{\mu/\alpha} \tilde D(\omega)\exp\left[-\frac{1}{\alpha}\sum_{\hat\tau,\hat\tau'=\hat\tau_\mathrm{min}}^{\hat\tau_\mathrm{max}} \tilde K(\omega,\tau)C^{-1}(\tau,\tau')\left(\int_0^\infty\frac{\dd\omega}{2\pi}\;\tilde\rho^*_{\alpha}(\omega)\tilde K(\omega,\tau')-\overline G(\tau')\right)\right],
\end{equation}
where we use the modified quantities, $\tilde \rho$, $\tilde D$ and $\tilde K$ introduced for SAI in Sec. \ref{Stochastic_Analytical_Inference} and
$\mu$ is a Lagrange multiplier to satisfy the normalization Eq.(\ref{normalization_of_field}) assuming $\tilde \rho$ and $\tilde D$ have the same normalization, i.e. $\int \frac{\dd \omega}{2\pi} \tilde \rho(\omega) = \int \frac{\dd \omega}{2\pi} \tilde D(\omega) = G(\tau_0)$.
On the other hand, in SAI $\chi^2[n]$ can be treated as the Hamiltonian for the system of the classical field $n(x)$. Thus, one can expand the Hamiltonian in the following way
\begin{equation}
\label{sai_expand_h}
\chi^2[n]=\int_0^{x_{max}}\ \dd x\ \epsilon(x)n(x)+\frac{1}{2}\int_0^{x_{max}}\ \dd x\dd y\ V(x,y)n(x)n(y)+\mathrm{const.},
\end{equation}
where we regard
\begin{equation}
\label{free_dispersion}
\epsilon(x)=-\sum_{\hat\tau,\hat\tau'=\hat\tau_\mathrm{min}}^{\hat\tau_\mathrm{max}}\overline G(\tau)C^{-1}(\tau,\tau') \tilde K(\phi^{-1}(x),\tau')
\end{equation}
as the free dispersion and 
\begin{equation}
\label{interaction_term}
V(x,y)=V(y,x)=\sum_{\hat\tau,\hat\tau'=\hat\tau_\mathrm{min}}^{\hat\tau_\mathrm{max}}\tilde K(\phi^{-1}(x),\tau)C^{-1}(\tau,\tau') \tilde K(\phi^{-1}(y),\tau')
\end{equation}
as the interaction. In the case of the mean field theory,  
\begin{equation}
\label{energy_form}
\chi^2[n]_{MF}=\int_0^{x_{max}}\ \dd x\ E(x)n(x)+\mathrm{const},
\end{equation}
where the energy of the system is obtained as
\begin{equation}
\label{energy}
E(x)=\left.\frac{\delta \chi^2[n]}{\delta n(x)}\right|_{n=\overline n}=\epsilon(x)+\int \dd y V(x,y)\overline n(y).
\end{equation}
With some effort one can work out the field configuration by using the saddle point method, which is
\begin{equation}
\label{saddle_point_solution}
\overline n(x)= e^{\mu/\alpha}\exp \left[-\frac{1}{\alpha}\left(\epsilon(x)+\int \dd y V(x,y)\overline n(y)\right)\right],
\end{equation}
where $\mu$ is again the Lagrange multiplier due to the normalization Eq.(\ref{normalization_of_field}).
So far one can see that actually Eq.(\ref{mem_image}) and Eq.(\ref{saddle_point_solution}) are equivalent.

Alternatively one can start from the aspect of entropy. One can consider a system consisting of indistinguishable particles in a canonical ensemble. Suppose that there are $M$ energy levels with degeneracies $m_p(p=1,2,...,M)$ and at each level there are $n_p$ particles. Then the number of equivalent microscopic occupancy configurations corresponding to a state specified by certain possible macroscopic field configuration $n(x)$ is $\Omega=\prod_p C_{n_p}^{m_p}$ and accordingly the entropy of this system can be written as
\begin{equation}
\label{S_one_configuration}
\begin{split}
S[n] &\equiv \ln \Omega[n]\\
&= \frac{1}{M}\sum_p \ln C_{n_p}^{m_p}\\
&\approx -\int \dd x\ n(x) \ln n(x),
\end{split}
\end{equation}
where in the third step we have used the Stirling's formula $\ln (m!)\approx m\ln m$ assuming $m_p\gg n_p$ and took the continuum limit $\frac{1}{M}\sum_p\rightarrow \int \dd x$, $m_p \rightarrow \infty$ and $n_p\rightarrow n(x)$. The entropy in SAI for all possible field configurations is
\begin{equation}
\label{sai_sai}
\begin{split}
S_{SAI} &\equiv \int \mathcal{D}n\ P[n]\ S[n]\\
&\approx \ln \Omega[\overline n]\\
&=-\int \dd x\ \overline n(x) \ln \overline n(x)\\
&=S_{MEM},
\end{split}
\end{equation}
where in the first step $P[n]$ is the probability of the system staying at configuration $n$ and the second step is obtained under the mean field approximation. We find that the implicit entropy in SAI is exactly the same to the one used in MEM. This verified the statement that MEM is the mean-field-limit of SAI~\cite{2004cond.mat..3055B}.

\subsubsection{SAI to SOM}
\label{SAI_to_SOM}
The $kink$ condition used in SOM is totally empirical. What SOM obtains is one special case in SAI. By setting $D(\omega)=K^{-1}(\omega,\tau_0)$ in Eq.(\ref{mapping}) one would arrive at $x=\omega/(2\pi)$ and $n(x)=\rho(\omega)K(\omega,\tau_0) = \tilde \rho(\omega)$. Using the $kink$ condition instead of averaging with $P[\alpha|D]$, we are able to obtain the possible solutions in SOM. One can infer that SAI with the default model $D(\omega)=K^{-1}(\omega,\tau_0)$ should give similar results to SOM. This is called the inverse kernel method in SAI. We will confirm this by model data tests given in Sec.\ref{Equivalence}.

\section{Implemention of SAI and SOM}
\label{Implemention}
\subsection{Monte Carlo evaluation for SAI} 
\begin{figure}[htb]
\centerline{\includegraphics[width=0.4\textwidth]{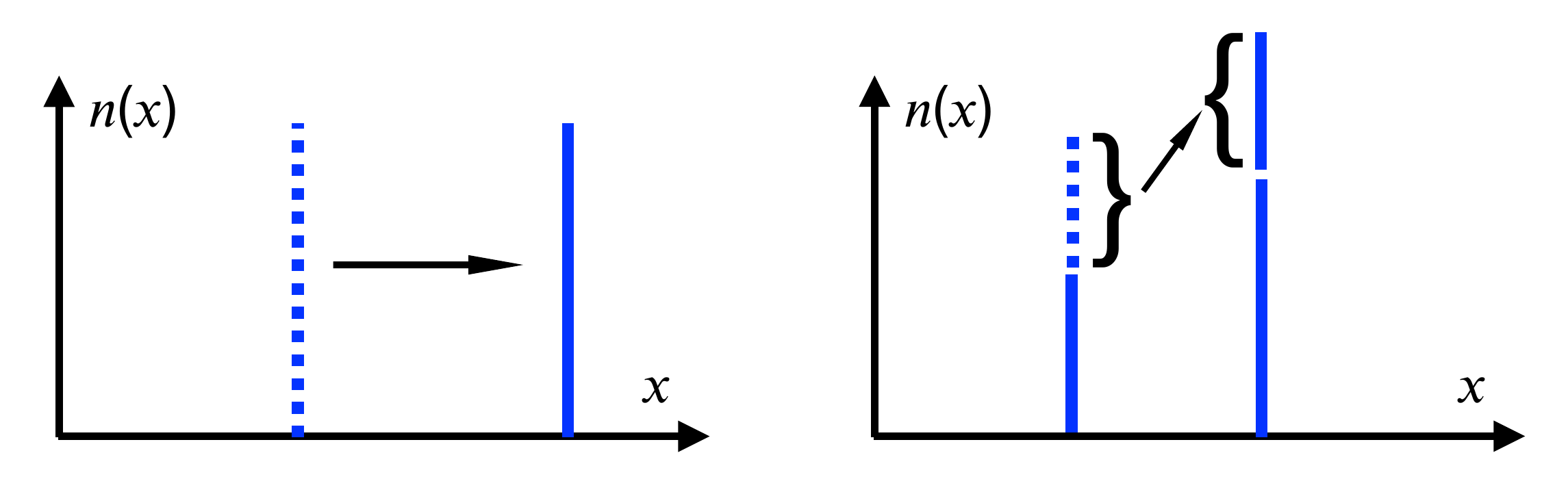} }
\caption{\textit{Left:} Shift a $\delta$-function. \textit{Right:} Residue sharing between two $\delta$-functions.}
\label{sai2}
\end{figure}

In this part we consider the Monte Carlo evaluation for SAI. The main work is to obtain $n(x)$ and $P[\alpha|D]$. Our procedures follow Ref.~\cite{2004cond.mat..3055B}. Firstly we represent $n(x)$ as a superposition of delta functions with residues $r_{\gamma}$ and position $a_{\gamma}$
\begin{equation}
\label{superposition}
n(x)=\sum_{\gamma} r_{\gamma}\ \delta(x-a_{\gamma}) \quad \text{with} \quad 0\leq a_{\gamma} \leq x_{max}.
\end{equation}
According to Eq.(\ref{normalization_of_field}) $n(x)$ needs to satisfy the normalization condition $\sum_\gamma\ r_{\gamma}=\overline{G}(\tau_0)$. Now we can perform two different kinds of updates to reshape the configuration $n(x)$ holding a detailed balance. The first one is to shift the position of a delta function
\begin{equation}
\label{update_1}
a_{\gamma} \mapsto a_{\gamma}'.
\end{equation}
The other one that dramatically improves the acceptance rate of attempted updates at low temperatures is the residue sharing in some subset $\Lambda$ of the delta functions
\begin{equation}
\label{update_2}
r_{\gamma} \mapsto r_{\gamma}'=r_{\gamma}+\sum_{\lambda \in \Lambda} \delta_{\gamma\lambda}\Delta r_{\lambda}
\end{equation}
that conserves higher moments
\begin{equation}
\label{higher_moments}
M^{(i)}=\int_0^{x_{max}} \dd x\ n(x)\ x^i=\sum_{\gamma}r^{\gamma}\ a_{\gamma}^i.
\end{equation}
To introduce such an update scheme let $\Lambda=\{\lambda_1,\lambda_2,...,\lambda_k\}=\{\lambda_1\}\cup\tilde{\Lambda}$, and we define a scale factor
\begin{equation}
\label{scale_factor}
Q_{\lambda}=\begin{cases}\ \ \ \ \ \ \ \ \ \ \ \ \ \ \ \ \ 1,\text{if}\ \lambda=\lambda_1\\ \frac{\prod \limits_{\mu \in {\tilde{\Lambda}}}(a_{\mu}-a_{\lambda_1})}{\prod  \limits_{\mu\in {\Lambda},\mu \neq \lambda}(a_{\mu}-a_{\lambda})},\text{if}\ \lambda \in \tilde{\Lambda} \end{cases},
\end{equation}
which satisfies $\sum_{\lambda=1}^{k} Q_\lambda a_\lambda^i=0$ for $i=0,1,\cdots,k-2$. Then, we can express the changes in residue as 
\begin{equation}
\label{updates_residues}
r_{\lambda}'=r_{\lambda}+\Delta r_{\lambda}=r_{\lambda}-sQ_{\lambda},
\end{equation}
where $s$ is randomly distributed in the interval
\begin{equation}
\label{s}
\max  \limits_{\lambda \in \Lambda^-}(r_{\lambda}/Q_{\lambda})<s<\min  \limits_{\lambda \in \Lambda^+}(r_{\lambda}/Q_{\lambda})
\end{equation}
to ensure the positivity of the residues, i.e. $r_{\lambda}'>0$. Here $\Lambda^-=\{\lambda\;|\; Q_{\lambda}<0\}$ and $\Lambda^+=\{\lambda\;|\; Q_{\lambda}>0\}$. In our study we randomly chose $k$ from 2 to 8 at each update. The updates mentioned above are schematically shown in Fig.\ref{sai2}. There can be an update changing the number of delta functions but we do not consider it in this study.

At each $\alpha$ configurations of $n(x)$ are generated with the Metropolis algorithm, where each update is accepted with a probability $P_\mathrm{adopt}=\min\{1,\exp (-\delta \chi^2/\alpha)\}$ where $\delta \chi^2$ is the difference of $\chi^2$ between two successive updates. We also used the parallel tempering~\cite{Marinari:1996dh} to obtain configurations at different temperatures simultaneously. The $\alpha$ range was divided into pieces with a constant ratio $\alpha_{i+1}/\alpha_{i}=R$. To represent a delta function we used a Gaussian function with a certain width, where the width was chosen in some range where the spectral function is stable.

\subsection{Monte Carlo evaluation for SOM}
\begin{figure}[htb]
\centerline{\includegraphics[height=0.2\textwidth, width=0.3\textwidth]{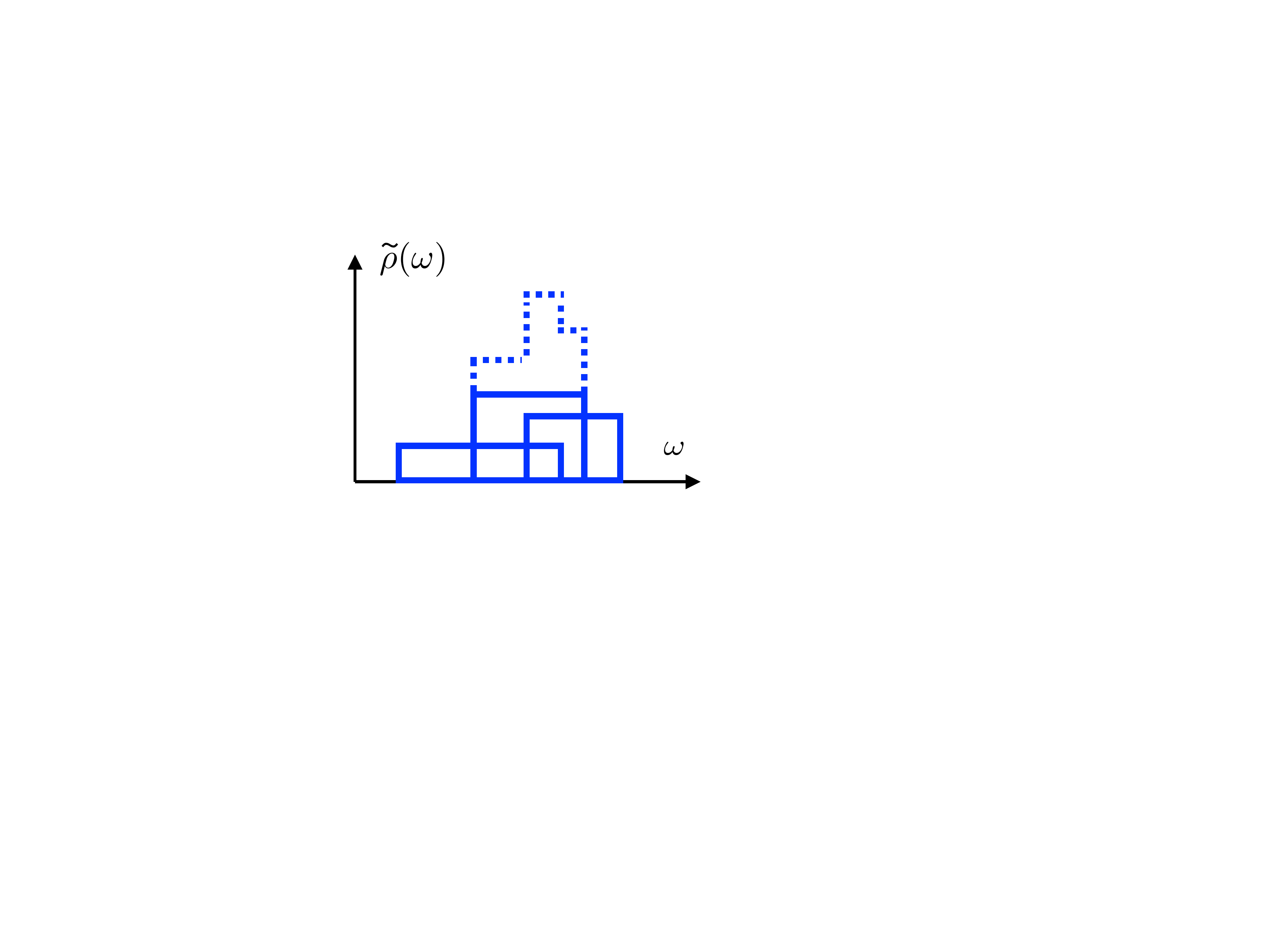} }
\caption{Spectral function constructed from overlapping boxes in SOM.}
\label{som1}
\end{figure}

In this section we discuss the Monte Carlo evaluation for SOM. We use the same basis as in Ref. \cite{PhysRevB.62.6317} where the spectral function $\rho(\omega)$ is parametrized as a sum of many boxes  
\begin{equation}
\label{simulated_spf}
\tilde \rho(\omega)=\sum_{t=1}^{K} \eta_{\{P_t\}}(\omega)
\end{equation}
with
\begin{equation}
\label{parameters_box}
 \eta_{\{P_t\}}(\omega)=\left\{
\begin{array}{rcl}
h_t, &	& \omega \in [c_t-w_t/2,c_t+w_t/2]\\ 
0, &		& \textit{otherwise,}
\end{array} \right. 
\end{equation}
where $w_t$, $h_t$, $c_t$ are width, height and center of a box, respectively. If two boxes overlap, the heights of the two boxes should be added up in the overlapping region as shown in Fig.\ref{som1} schematically.
The normalization condition for boxes is as follows:
\begin{equation}
\label{normalization}
\int_0^{\omega_{max}}\ \frac{d \omega}{2\pi}\ \tilde \rho(\omega)=\sum_{t=1}^K h_t w_t = \overline G(\tau_0).
\end{equation} 

The elementary updates can be realized by changing a random parameter of the boxes in the sets $\{{P_t}\}=\{h_t,w_t,c_t\}$. During the updates the number of the boxes and the sum of their area are fixed. And the change of parameters must sit in the domains of definitions of a box $\Xi$, which are ${h_t} \in [h_{min},h_{max}]$, ${w_t} \in [w_{min},w_{max}]$ and ${c_t} \in [\omega_{min},\omega_{max}]$. The elementary updates used in SOM are listed as follows and depicted in Fig.\ref{som2}:
 
\begin{figure}[htb]
\centerline{\includegraphics[height=0.2\textwidth, width=0.8\textwidth]{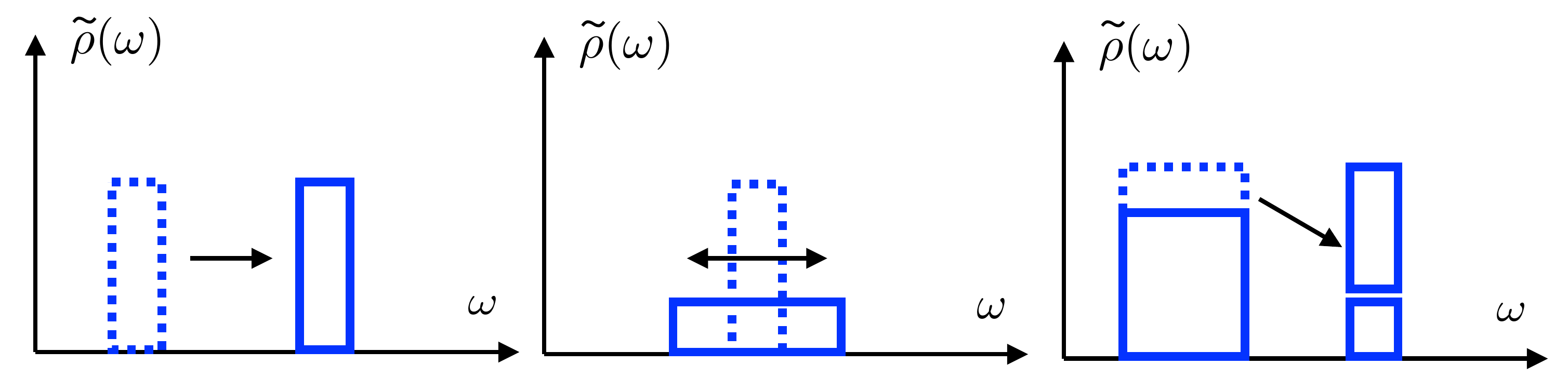} }
\caption{\textit{Left:} Shift a box. \textit{Middle:} Change a box. \textit{Right:} Share height between two boxes.}
\label{som2}
\end{figure}

(I) \emph{Shift a box.} Vary the center $c_t$ of a randomly selected box stochastically. Newly generated $c_t$ is restricted to be in the domain of definition $\Xi_{ct}=[\omega_{min},\omega_{max}]$.

(II) \emph{Change a box.}  Change the height of a randomly chosen box from $h_t$ to $h_t'$  keeping the center $c_t$ and area fixed. The width is subsequently changed from $w_t$ to $w_t\cdot h_t/h_t'$. Newly generated $h_t'$ and $w_t'$ are restricted in the domains $\Xi_{ht}=[h_{min},h_{max}]$ and $\Xi_{wt}=[w_{min},w_{max}]$.
 
(III) \emph{Share height between two boxes.} Choose two boxes A and B randomly. Cut part of the height of A and add this part to B. The centers of box A and box B are fixed. The sum of the area of box A and box B is fixed, too. In this update the height of box A is changed from $h_{tA}$ to $h_{tA}'$ and the height of box B is changed from $h_{tB}$ to $h_{tB}'=h_{tB}+w_{tA}\cdot (h_{tA}-h_{tA}')/w_{tB}$. Newly generated $h_{tA}'$ and $h_{tB}'$ are restricted by $\Xi_{ht}=[h_{min},h_{max}]$.  

(IV) \emph{Share width between two boxes.} Same as (III) but in this update we change $w_t$ instead of $h_t$. The aim of using update (III) and (IV) is to make a connection between two boxes helping to reshape the boxes more efficiently.

Similarly to SAI configurations of boxes are generated by the Metropolis algorithm. Except that the basis which we used here is the same as in Ref. \cite{PhysRevB.62.6317}, the probability $P_\mathrm{adopt}$, the types of elementary updates and how the final solution is obtained are quite different. The original SOM has more types of elementary updates than what we used in this work, for instance add/remove/split a box and glue two boxes. Since these updates can be obtained from the combinations of the four elementary updates mentioned above and do not show any advantage, we use the current types of updates instead. 

In principle,  different choices of basis for parametrization of the spectral function in SAI and SOM are equivalent. Our choices of rectangle and delta function basis for SOM and SAI, respectively, are just a matter of convenience. In particular, the choice of delta function basis for SAI makes the $x\rightarrow \omega$ mapping quite simple during the practical implementation.

\section{Analyses with model data }
\label{mock_data_test}
In this section we show the results from the model data tests using MEM, SOM and SAI.
We will firstly show model spectral functions used in the tests in Sec.~\ref{mockspf}, and then illustrate the equivalence of SAI and SOM numerically in Sec.~\ref{Equivalence}. We will discuss the dependencies of output spectral function
on the noise level $\epsilon$ and the number of data points $N_{\tau}$ in Sec.~\ref{dependence_noise_N} as well as on the default model 
in Sec.~\ref{dependence_on_DM}.

% Finally we apply our methods in real lattice data taken from Ref.\cite{Ding:2012sp} in Sec.\ref{real_data_analysis} . 

 Note that in the following model data tests we use dimensionless quantities. For instance, the dimensionless frequency $\hat{\omega}$ is related to the dimensional one through $\hat{\omega}=a\omega$ where $a$ is the lattice spacing. We also use the conventions $\hat{\rho}=a^2\rho$ and $\hat{T}=aT$ in the model data tests.

%
%We will show mock spe
%I we show. SAI results based on different averaging criteria are also discussed. In Sec.\ref{dependence_noise_N} we show the dependence on noise level $\epsilon$ and number of data points $N_{\tau}$.  Finally we summarize in Sec.\ref{conclusion}.

\subsection{Model spectral functions}
\label{mockspf}
The model spectral functions are constructed considering two different physics inspired cases. In the tests we mainly focus on the two cases:

(A) $\hat{\rho}_{below}(\hat{\omega})=\widetilde \Theta(\hat{\omega},\hat{\omega}_1,\Delta_1)(1-\widetilde \Theta(\hat{\omega},\hat{\omega}_2,\Delta_2))\hat{\rho}_{res}(c_{res1},\Gamma_1,M_1)+\widetilde \Theta(\hat{\omega},\hat{\omega}_3,\Delta_3)\hat{\rho}_{cont}$ corresponding to the spectral function at a temperature below $T_c$. Here $\hat{\rho}_{res}$ denotes a resonance peak and $\hat{\rho}_{cont}$ denotes a free continuum part.

(B) $\hat{\rho}_{above}(\hat{\omega})=\hat{\rho}_{trans}+\widetilde \Theta(\hat{\omega},\hat{\omega}_4,\Delta_4)(1-\widetilde \Theta(\hat{\omega},\hat{\omega}_5,\Delta_5))\hat{\rho}_{res}(c_{res2},\Gamma_2,M_2)+\widetilde \Theta(\hat{\omega},\hat{\omega}_6,\Delta_6)\hat{\rho}_{Wilson}$ corresponding to the spectral function at a temperature above $T_c$. Here $\hat{\rho}_{trans}$ denotes a transport peak, and a free Wilson spectral function denoted as $\hat{\rho}_{Wilson}$ is also introduced to take into account the lattice cutoff effects.

The elementary parts of the model spectral functions needed in $\hat{\rho}_{below}(\hat{\omega})$ and $\hat{\rho}_{above}(\hat{\omega})$ are smoothed by a modified $\Theta$-function $\widetilde \Theta(\hat{\omega},\hat{\omega}_{i},\Delta_{i})=\big(1+\exp(\frac{\hat{\omega}_{i}^2-\hat{\omega}^2}{\hat{\omega}\Delta_{i}})\big)^{-1}$ to make the model spectral functions more realistic. The elementary parts of the spectral functions are listed below:

\begin{enumerate}
\item Transport peak
\begin{align}
\label{tansport_peak}
\hat{\rho}_{trans}(c_{trans},\eta)=c_{trans}\,\frac{\hat{\omega} \eta}{\hat{\omega}^2+\eta^2}.
\end{align} 
\item Resonance peak
\begin{align}
\label{resonance_peak}
\hat{\rho}_{res}(c_{res},\Gamma,M)=c_{res}\,\frac{\Gamma M \hat{\omega}^2}{(\hat{\omega}^2-M^2)^2+M^2\Gamma^2}.
\end{align}
\item Free continuum spectral function
\begin{equation}
\label{continuum_peak}
\begin{split}
\hat{\rho}_{cont}(c_{cont},m_c)&=c_{cont}\,\frac{N_c}{8\pi}\,\Theta(\hat{\omega}^2-4m_c^2)\,\hat{\omega}^2 \tanh \left(\frac{\hat{\omega} N_{\tau}}{4}\right)\\
&\times \sqrt{1-\left(\frac{2m_c}{\hat{\omega}}\right)^2}\, \left[a^{(1)}+a^{(2)}(\frac{2m_c}{\hat{\omega}})^2\right].
\end{split}
\end{equation}
\item Free Wilson spectral function
\begin{equation}
\label{wilson_peak}
\begin{split}
\hat{\rho}_{Wilson}(c_{Wilson},m)&=c_{Wilson}\frac{4\pi N_c}{{N_{\sigma}}^3} \sum_{\emph{\textbf k}}\sinh (\frac{\hat{\omega}}{2\hat{T}})\left[b^{(1)}-b^{(2)}\frac{\sum_{i=1}^{3} \sin^2 k_i}{\sinh^2 E_{\emph{\textbf k}}(m)}\right]\\
&\times \frac{\delta(\hat{\omega}-2E_{\emph{\textbf k}}(m))}{2\left(1+\emph {M}_{\emph{\textbf k}}(m)\right)^2 \cosh^2 \left(\frac {E_{\emph{\textbf k}}(m)}{2\hat{T}}\right)},
\end{split}
\end{equation}
where
\begin{equation}
\begin{split}
\label{cosh}
&b^{(1)}=\frac{a^{(1)}-a^{(2)}}{2} ,\,\,b^{(2)}=\frac{a^{(2)}-a^{(3)}}{2}, ~~~\cosh E_{\emph{\textbf k}}(m)=1+\frac{ K^2_{\emph{\textbf k}}+M^2_{\emph{\textbf k}}(m)}{2(1+M_{\emph{\textbf k}}(m))},\\
&K_{\emph{\textbf k}}=\sum_{i=1}^{3} \gamma_i \sin k_i, ~~~\emph{M}_{\emph{\textbf k}}(m)=\sum_{i=1}^{3}(1-\cos k_i)+m.
\end{split}
\end{equation}
\end{enumerate}

\begin{table}[htb]
\begin{center}
\begin{tabular}{|c|c|}
		\hline
		Spectral function & Parameters \\ \hline
		$\hat{\rho}_{res1}$ \ & $c_{res1}=0.08/7$, $\Gamma_1=0.05$, $M_1=0.155$\\ 
		$\hat{\rho}_{cont}$ \ & $c_{cont}=1$, $a^{(1)}=2$, $a^{(2)}=1$, $m_c=0.0775$, $N_c=3$\\
                $\hat{\rho}_{trans}$ \ & $c_{trans}=5\times 10^{-5}$, $\eta=0.006$\\
		$\hat{\rho}_{res2}$ \ & $c_{res2}=0.06$, $\Gamma_2=0.15$, $M_2=0.225$\\ 
		$\hat{\rho}_{Wilson}$ \ & $c_{Wilson}=1$, $b^{(1)}=3$, $b^{(2)}=1$, $m=0.073$, $N_c=3$, $N_{\tau}=48$, $N_{\sigma}=2304$\\ \hline
	\end{tabular}
\end{center}
\caption{Parameters for the model spectral functions.}
\label{Mock_SPFs_Parameters}
\end{table}

\begin{table}[htb]
\begin{center}
\begin{tabular}{|c|c|}
		\hline
		$\hat{\omega}_1=0.145$ \ & $\Delta_1=0.01$\\ 
                $\hat{\omega}_2=0.155$ \ & $\Delta_2=0.05$\\ 
                $\hat{\omega}_3=0.225$ \ & $\Delta_3=0.05$\\  
                $\hat{\omega}_4=0.225$ \ & $\Delta_4=0.15$\\ 
                $\hat{\omega}_5=0.225$ \ & $\Delta_5=0.15$\\ 
                $\hat{\omega}_6=0.350$ \ & $\Delta_6=0.2$\\\hline
	\end{tabular}
\end{center}
\caption{Parameters for $\widetilde \Theta$-functions.}
\label{theta_Parameters}
\end{table}

The parameters used in each part of the model spectral functions are summarized in Table~\ref{Mock_SPFs_Parameters}. They are chosen to mimic physical situations given the lattice spacing $a^{-1}=$20 GeV. At temperatures below $T_c$, the resonance peak has the mass of $J/\psi$ meson ($\sim$3.1 GeV). The spectral function at $T>T_c$ has a transport peak and the resonance peak might disappear and a broader peak should appear at larger energy. Accordingly we shift the resonance peak from $M_1=0.155$ to $M_2=0.225$. The transport peak, which is expected to be a Breit-Wigner like distribution, corresponds to $2\pi TD\sim2$ with $\chi_{00}/T^2\sim0.07$ where $D$ is the quark diffusion coefficient and $\chi_{00}$ is the quark number susceptibility. In the free continuum spectral function and free Wilson spectral function the mass of quark is set to be $\sim$1.5 GeV, and the threshold of these free spectral function can be modified by using the modified $\Theta$-function in the bound state region. The parameters in the modified $\Theta$-function used are listed in Table~\ref{theta_Parameters}.

With the model spectral functions given above, the model correlators are generated by adding a Gaussian noise with a standard deviation $\sigma=\epsilon\cdot\bar{G}\cdot\tau$ where $\epsilon$ is the noise level. In our model data tests $\hat \tau_0$ is always set to 1. We also set $\hat \tau_\mathrm{min}$ and $\hat \tau_\mathrm{max}$ to 1 and $N_\tau/2$, respectively, where $N_\tau$ is the temporal lattice size.

%\clearpage{}

\subsection{Equivalence of SAI and SOM}
\label{Equivalence}

\begin{figure}[htpb]
\centerline{\includegraphics[ width=0.8\textwidth]{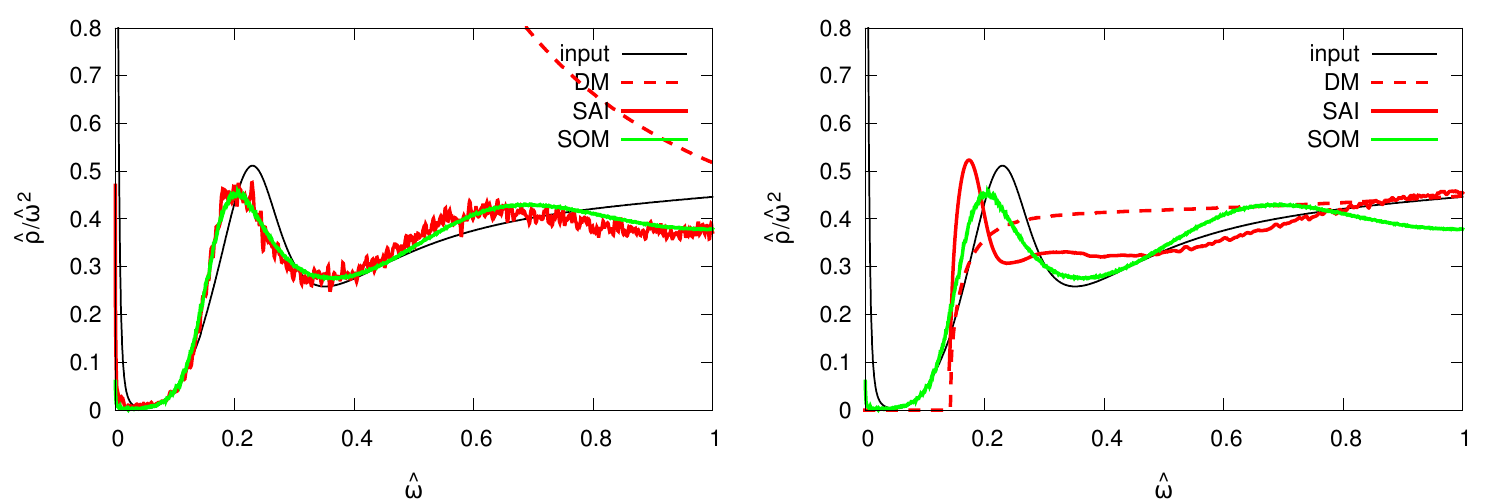}}
%\centerline{\includegraphics[height=0.3\textwidth, width=1\textwidth]{free_w.pdf}}
\caption{A comparison of spectral functions obtained from SOM and SAI using DM $K^{-1}(\hat{\omega})$ (\textit{left}) and a rescaled free Wilson spectral function (\textit{right}) as the default model. The black solid curve is the input model spectral function, and the red dashed curve denotes the default model. The red and green solid curves represent the output spectral functions obtained using the SAI and SOM, respectively. }
\label{somvssai}    
\end{figure}

As discussed at the end of Sec.~\ref{SAI_to_SOM}, SAI is equivalent to SOM given the default model $D(\hat{\omega})=K^{-1}(\hat{\omega},\tau_0)$. In this section we will show the equivalence numerically using model correlator data with $N_{\tau}=48$ and $\epsilon=5\times10^{-3}$. The correlators are computed using $\hat{\rho}_{above}(\hat{\omega})$ as shown in Sec.~\ref{mockspf}.
As seen from the left panel of Fig.~\ref{somvssai}, where the default model $D(\hat{\omega})=K^{-1}(\hat{\omega})$ is used in the SAI analyses, the output spectral function obtained using the SAI is almost the same as that obtained using the SOM. 
For comparison output spectral function obtained from the SAI using a default model different from $K^{-1}(\hat{\omega},\tau_0)$, i.e. 
a rescaled free Wilson spectral function is also shown in the right panel of Fig.~\ref{somvssai}. It is clearly seen that the obtained spectral function is different from that obtained using SOM.

The discussion above verifies the statement that SOM is just one special case of SAI, and they should give similar results when $D(\hat{\omega})=K^{-1}(\hat{\omega},\tau_0)$ is used in SAI.

\subsection{ Dependences on $N_{\tau}$ and noise level $\epsilon$}
\label{dependence_noise_N}

\begin{figure}[htb]
\centerline{\includegraphics[width=0.95\textwidth]{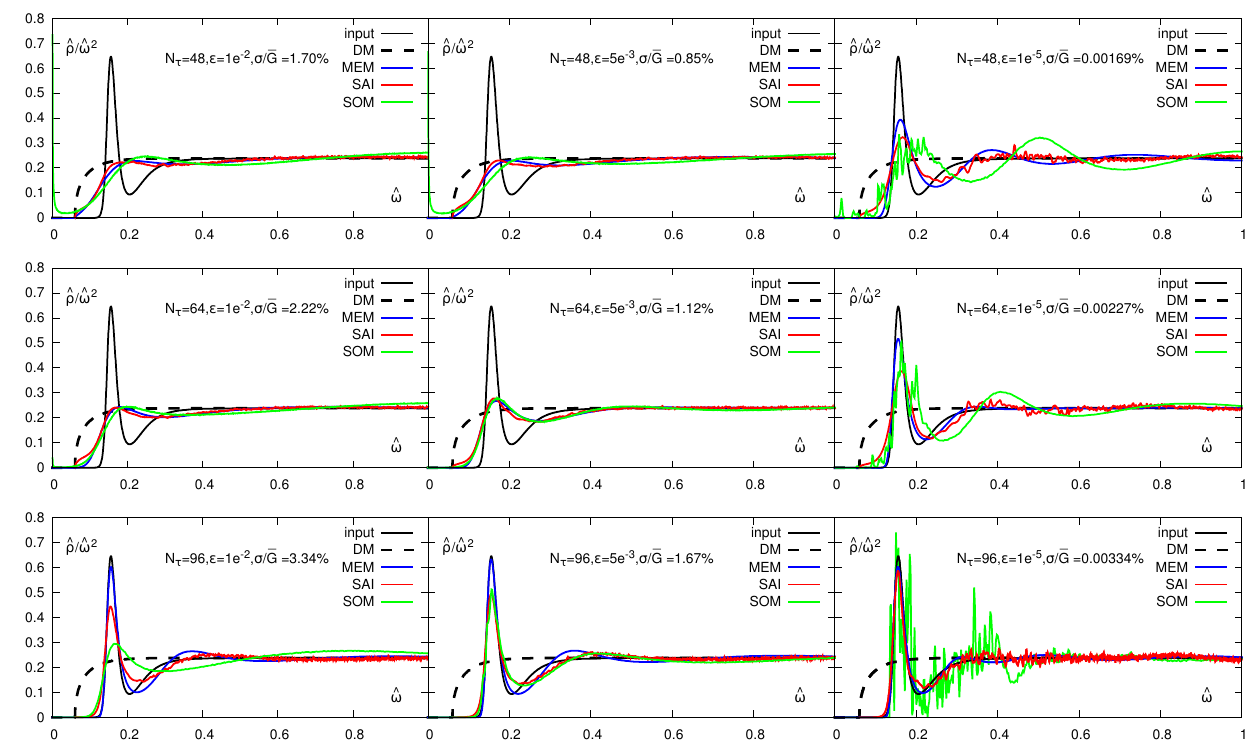} }
\caption{Dependence on $N_{\tau}$ and noise level $\epsilon$ of output spectral functions obtained from SOM, SAI and MEM. The black solid curve is the input spectral function, and the black dashed curve denotes the DM. The other colored curves are output spectral functions. From \textit{top} to \textit{bottom} $N_{\tau}$=48, 64 and 96. From \textit{left} to \textit{right} $\epsilon=10^{-2},~5\times10^{-3}$ and $10^{-5}$.}
\label{Dependence_N_noise}
\end{figure}

The number of data points in the correlators and the quality of the data have crucial influence on the reconstructed output spectral functions. To show this we analyze the model data with $N_{\tau}=48,64,96$ and $\epsilon=10^{-5},5\times10^{-3},10^{-2}$, in which $\epsilon=5\times10^{-3}$ is close to the state-of-the-art quality of data obtained from lattice QCD simulations. In this test we choose $\hat{\rho}_{below}(\hat{\omega})$ as the input model spectral function, and in SAI and MEM we use a free continuum spectral function as the DM. This DM has a similar behavior as the input spectral function in the large $\hat{\omega}$ part. Note that both in the input spectral function and the default model a transport peak was not introduced. The results are summarized in Fig.\ref{Dependence_N_noise}. We plot $\hat{\rho}(\hat{\omega})/\hat{\omega}^2$ as a function of $\hat{\omega}$ to suppress the rise of spectral functions in a very large energy range. The range of $\hat{\omega}$ used in the analyses of these three approaches is $[0,4]$ but for a better illustration we only show the results in $[0,1]$ in the figure. For a better illustration the ratio of the standard error to the mean values of the correlators (denoted by $\sigma/\bar G$) at the middle point ($\tau=N_\tau/2$) of the correlators is also given.

From Fig.~\ref{Dependence_N_noise} we can see that in all cases the free continuum part can be reproduced very well while the resonance part strongly depends on $N_{\tau}$ and $\epsilon$. When the data are noisy or the number of data points is not sufficiently large, i.e. in the case of $N_{\tau}=48,\epsilon=10^{-2},5\times10^{-3}$(top-left and top-middle) and $N_{\tau}=64,\epsilon=10^{-2}$(middle-left), all three methods can only give a rough structure of the resonance, and SOM even gives fake transport peak in the case of $N_{\tau}=48,\epsilon=10^{-2},5\times10^{-3}$ (top-left and top-middle panels). In the case of a larger $N_{\tau}$ and a smaller noise-to-signal ratio it is expected to see that the fake transport peak obtained from SOM starts to disappear, and the output resonance peaks obtained from all three methods approach to the input one. 

\begin{figure}[h]
\centerline{\includegraphics[width=0.8\textwidth]{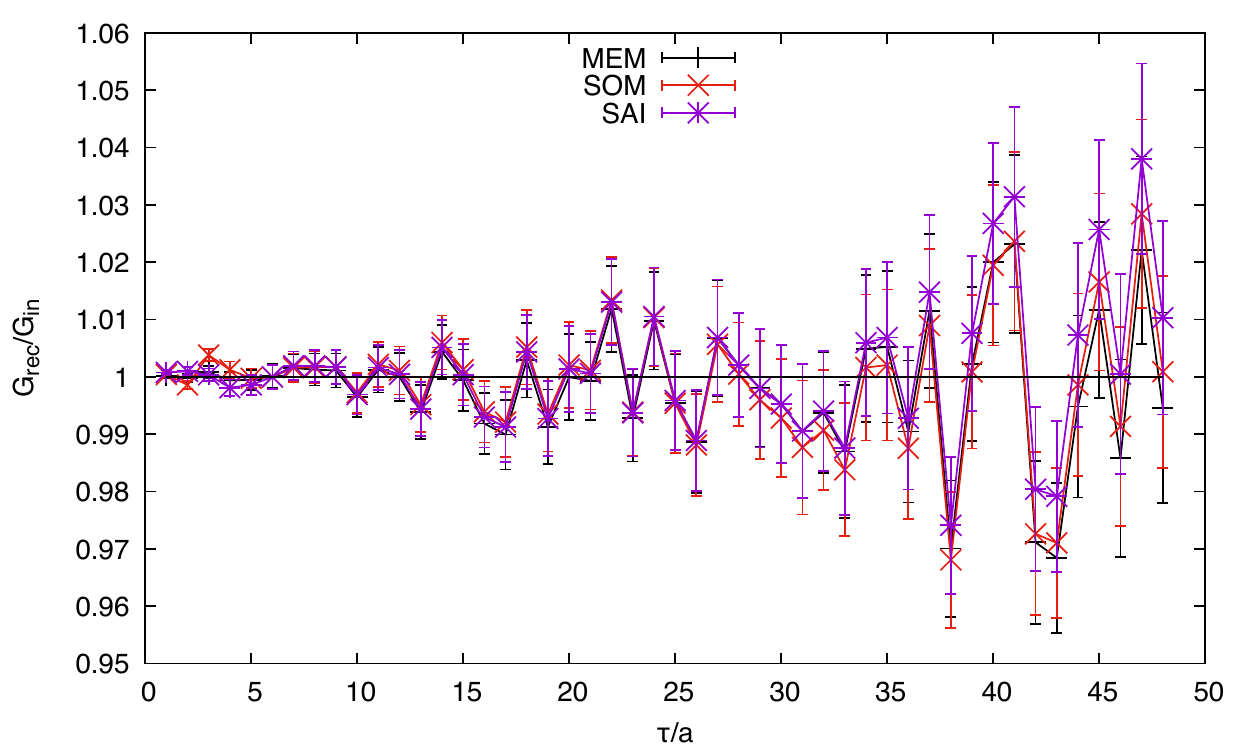} }
\caption{The ratio of the correlators reconstructed from the output spectral functions obtained by three methods to the input ones at $N_{\tau}=96, \epsilon=5\times 10^{-3}$. The error bars shown in the figure are only from the input correlators. }
\label{Gout_Gin}
\end{figure}

At $N_{\tau}=96, \epsilon=5\times 10^{-3}$, we examine the ratio of the correlators reconstructed from the output spectral functions obtained by the three methods to the input ones. The results are shown in Fig.\ref{Gout_Gin}. We found that all three methods give consistent results, just like the spectral functions themselves, and the ratios are close to unity at all the distances.

In the case of a very small noise-to-signal ratio, e.g. at the middle point $\sigma/\bar{G}=0.00334$\% (see bottom-right panel of Fig.~\ref{Dependence_N_noise}), the output spectral functions obtained from the stochastic methods, i.e. SOM and SAI, show some small wiggles in the large $\hat{\omega}$ region. The wiggling behavior even shows up in the smaller energy region in the spectral function obtained from SOM.
%
%, the peak-location of the resonance cannot be reproduced well from SOM.
%The exception is that for SAI and especially SOM, which even cannot reproduce the peak-location right, when the noise level is quite small(see the right panel of Fig.\ref{Dependence_N_noise}), the small $\hat{\omega}$ parts reconstructed have strong fluctuations. 
The reason may be that MEM works in a deterministic way of solving the equations[see Eq.~(\ref{direvative_of_q})], i.e. leaves the smoothness of the default model, while the SOM and SAI are of a stochastic nature. At some quite small noise level the current limited number of stochastic samplings cannot reflect the noise level precisely, and the situation can be improved with larger number of samplings.

In the realistic lattice QCD simulations the noise-to-signal ratio at the middle point is much larger than 0.00334\% and is similar to the noise level shown in the left and middle panels of Fig.~\ref{Dependence_N_noise}. Among these nine figures the quality of the data shown in the bottom-middle panel, i.e. with $N_\tau=96$ and $\sigma/\bar{G}=1.67$\%, is most similar to the state-of-the-art lattice QCD simulations for temporal correlation functions at $T<T_c$. We can see that in this case all three methods succeed in reconstructing the general peak structure of the resonance peak as well as the continuum part of the input spectral function. And even with a 2-times-larger noise level as shown in the left panel with $N_\tau=96$ the peak location of the resonance peak is always reproduced well using the structureless free continuum spectral functions as the default model. For the reconstruction of the peak height and the width of the peak MEM seems to be better than the SOM and SAI which tend to give a larger width and a smaller peak height. It needs to be noted that the current output spectral functions are obtained using only one and a simple default model, and in the next section we will discuss the dependence of the reconstructed spectral functions on default models.

%\clearpage{}

\subsection{ Dependence on default model}
\label{dependence_on_DM}
In this section we study the dependence of output spectral functions on default models at temperatures both below and above $T_c$. In each case we consider only one model spectral function and try to reconstruct it with various DMs. At $T<T_c$  the model correlators are produced using the spectral function $\hat{\rho}_{below}(\hat{\omega})$ with $N_{\tau}=96$ and a noise level $\epsilon=2.5\times10^{-3}$, while at  $T>T_c$  the model correlators are produced using the spectral function $\hat{\rho}_{above}(\hat{\omega})$ with $N_{\tau}=48$ and a noise level $\epsilon=5\times10^{-3}$. These noise levels are chosen to mimic the case in the real lattice data. The main differences between the spectral functions at $T<T_c$ and $T>T_c$ in our current model data tests are 1) there is no transport peak in $\hat{\rho}(T<T_c)$, and there exists one in $\hat{\rho}(T>T_c)$ and 2) the resonance peak in $\hat{\rho}(T>T_c)$ is located at a larger value of $\hat{\omega}$ and has a broader width than that in $\hat{\rho}(T<T_c)$.

\subsubsection{Default model dependence of $\hat{\rho}(T<T_c)$}
\label{Below_$T_c$}

First we consider the case at temperatures below $T_c$. In this case we use four different default models for SAI and MEM analyses. $DM1$ is simply a rescaled free continuum spectral function. $DM2$ has an additional transport peak to $DM1$. $DM3$ and $DM4$ are of the same type as the input spectral function but $DM3$ has a smaller resonance peak location than the input spectral function while $DM4$ has a larger one. For convenience hereafter we suppress all the normalization factors coming from the normalization condition as seen from Eq.~(\ref{normalization_of_field}) in the notation of the default models. The parameters used in these default models are listed in Table~\ref{DM_Parameters0}.

\begin{table}[htb]
\begin{center}
\begin{tabular}{|c|c|c|}
		\hline
		Default model \ & Type \ & Parameters\\		\hline
		DM1 \ & $\hat{\rho}_{cont}$ \ & $m_c=0.03$ \\ 		\hline
		DM2 \ & $\hat{\rho}_{trans}+\rho_{cont}$ \ & $m_c=0.03$ \\ 		\hline
		DM3 \ & $\hat{\rho}_{below}$ \ & $M=\hat{\omega}_1=\hat{\omega}_2=0.1$ \\  		\hline
		DM4 \ & $\hat{\rho}_{below}$ \ & $M=\hat{\omega}_1=\hat{\omega}_2=0.225,\Delta_1=\Delta_2=0.1$ \\ \hline
	\end{tabular}
\end{center}
\caption{Parameters of the default models at temperatures below $T_c$. }
\label{DM_Parameters0}
\end{table}

\begin{figure}[htbp]
\centerline{\includegraphics [width=0.8\textwidth]{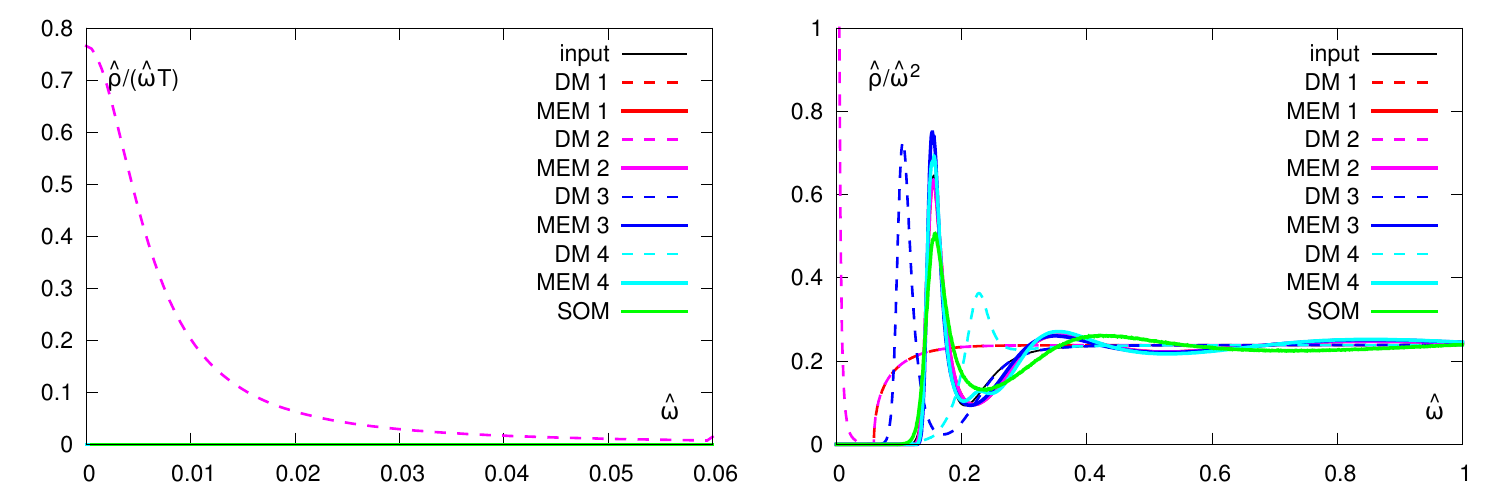} }
\centerline{\includegraphics[width=0.8\textwidth]{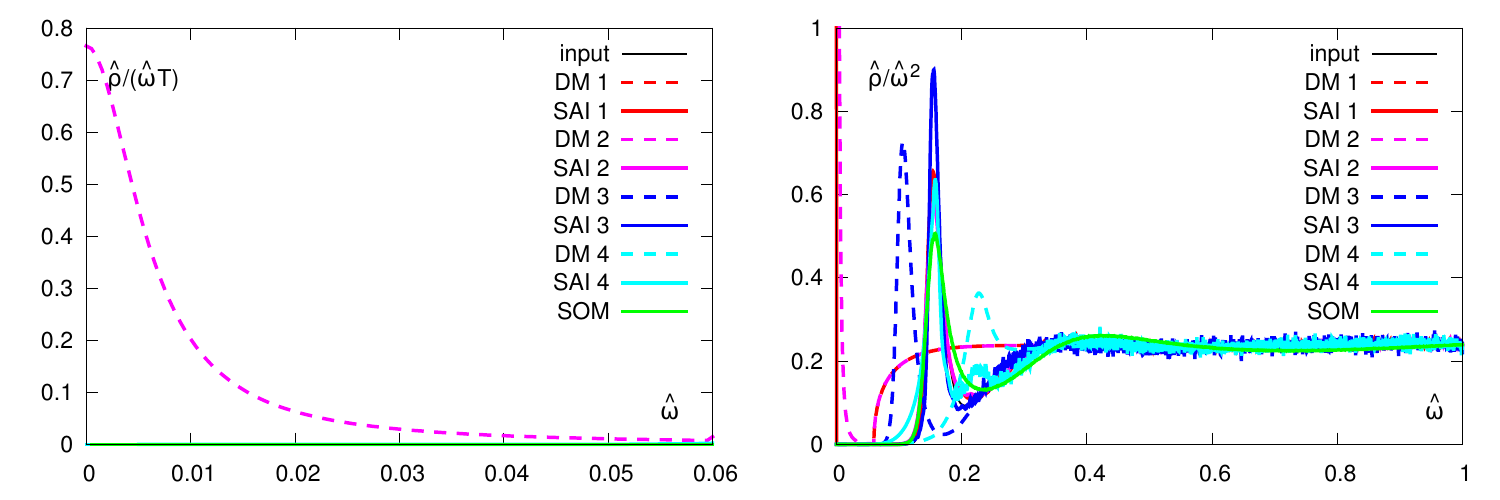} }
\caption{Default model dependencies of $\hat{\rho}(T<T_c)$ obtained from SAI (bottom panel) and MEM (top panel). SOM results are also shown with green solid curves. The left panels show the spectral functions in the small energy region while the right ones show the spectral function in the whole energy region. The model correlators are produced using $\hat{\rho}_{below}(\hat{\omega})$ with $N_{\tau}=96$ and a noise level $\epsilon=2.5\times10^{-3}$. }
\label{dm_below}
\end{figure}

The spectral functions in the small energy region given by MEM, SOM and SAI analyses are shown in the
 left panels of Fig.~\ref{dm_below}. The conclusion is the same as obtained from Sec.~\ref{dependence_noise_N}:
 all three methods give reliable results on the small energy region of the spectral function if there is no transport peak in the input spectral function.

% It can be seen that the output spectral function obtained from SOM is comparable to zero in this region, which is exactly the situation in the input spectral function. And in the case of MEM and SAI, the output spectral functions are also vanishing even though a transport peak is put in the default model, i.e. $DM2$. These findings shows the reliability of these three methods in reproducing the small energy region if there is no transport peak in the input spectral functions.   

%
% The results are summarized in Fig.~\ref{dm_below}. The top and bottom panels show the results obtained from MEM and SAI, respectively.  The conclusion 
% 
% 
% For comparison the results of SOM are included in both panels.

%
%
%are exactly 0 for all default models while in SAI it is $1.76\times 10^{-05}$, $1.72\times 10^{-04}$, $8.89\times 10^{-92}$ for $DM1-DM4$. We conclude that at $N_{\tau}=96,\epsilon=2.5\times10^{-3}$, if there is no transport peak in the input model SPF, one can find no/compatible with 0 transport peak appears in the outputs when using SOM$\&$MEM$\&$SAI analysis, even if it was given in the DM. 

The spectral functions in the whole energy region are shown in the right panels of Fig.~\ref{dm_below}. It is found that in the MEM analysis the peak locations of the reconstructed resonance peak obtained using $DM1-DM3$ are $\hat \omega=$0.1550, 0.1550 and 0.1530 while SAI analysis gives the peak locations at $\hat \omega=$0.1551, 0.1545 and 0.1558. Thus the default model dependence of the reconstructed peak location is very small. In comparison, the SOM analysis shows a peak location at $\hat \omega=$0.1575. We can see that the reconstructed peak locations by all three methods are very close to the input one $M_1=$0.155. 
%One should also notice that the output spectral function based on $DM4$ has an extra fake %resonance peak, which is not the case in the analyses using other default models and also %in the input spectral function. This suggests that only the existence of an first peak %structure is reliable in the current model data tests.
%We conclude that in this case when given different DMs, SAI and MEM almost have very mild %DM dependence on the peak-location of the first resonance peak and can reproduce the peak %location of the input spectral function. The other aspect about the resonance peak is its %height and width. 
As seen from Fig.~\ref{dm_below} the peak height and width obtained by all three methods obviously differ from those of the input spectral function and have a relatively larger default model dependence. Thus the information on the peak height and width extracted from these methods are not as reliable as the peak location.

\subsubsection{Default model dependence of $\hat{\rho}(T>T_c)$}

In these tests we consider the case at a temperature above $T_c$.
A big difference in the model spectral function of $\hat{\rho}(T>T_c)$ from $\hat{\rho}(T<T_c)$ is that there is an additional transport peak.
Thus we want to test the default model dependence of the output spectral function by varying the low frequency and high frequency
part of the DM separately.  In this case we will use eight different default models for analysis listed as follows.
\begin{itemize}
\item  $DM1$ and $DM2$ are composed of only rescaled free Wilson spectral functions. The difference between $DM1$ and $DM2$ is the threshold of the free Wilson spectral function, i.e. different values of quark masses.
% $DM3-DM8$ have the same type as the input model SPF $\hat{\rho}_{above}(\hat{\omega})$.

 \item $DM3$ and $DM4$ have similar transport peaks to the input spectral function,  but the resonance peak in the $DM3$ has a smaller peak location than that in the input spectral function while the one in the $DM4$ has a larger peak location.
 
  \item $DM5$, $DM6$ and $DM7$ have resonance peaks which have the same peak location as the input spectral function. And the width of the transport peak is also same as the input spectral function but the heights of the transport peak are different from the input one and among each other.
 
 \item $DM8$ has the same resonance peak location and the same transport peak-height ($\varpropto c_{trans}/\eta$) as $DM6$ but has a different width of the transport peak. 
 \end{itemize}
 
\begin{figure}[h]
\centerline{\includegraphics[width=0.8\textwidth]{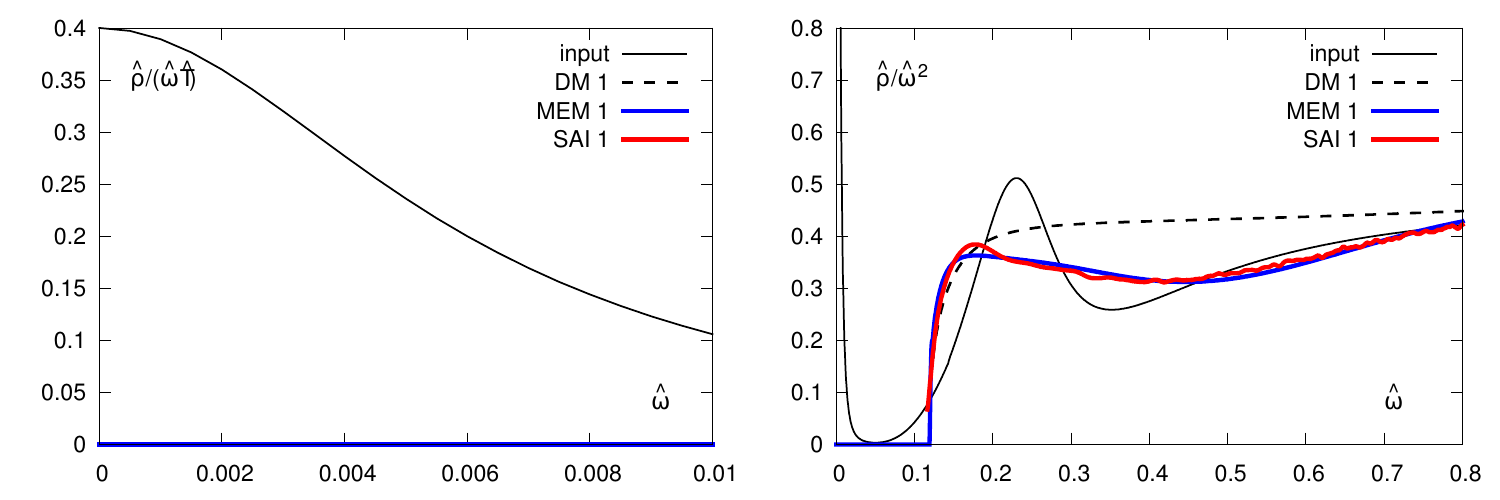} }
\centerline{\includegraphics[width=0.8\textwidth]{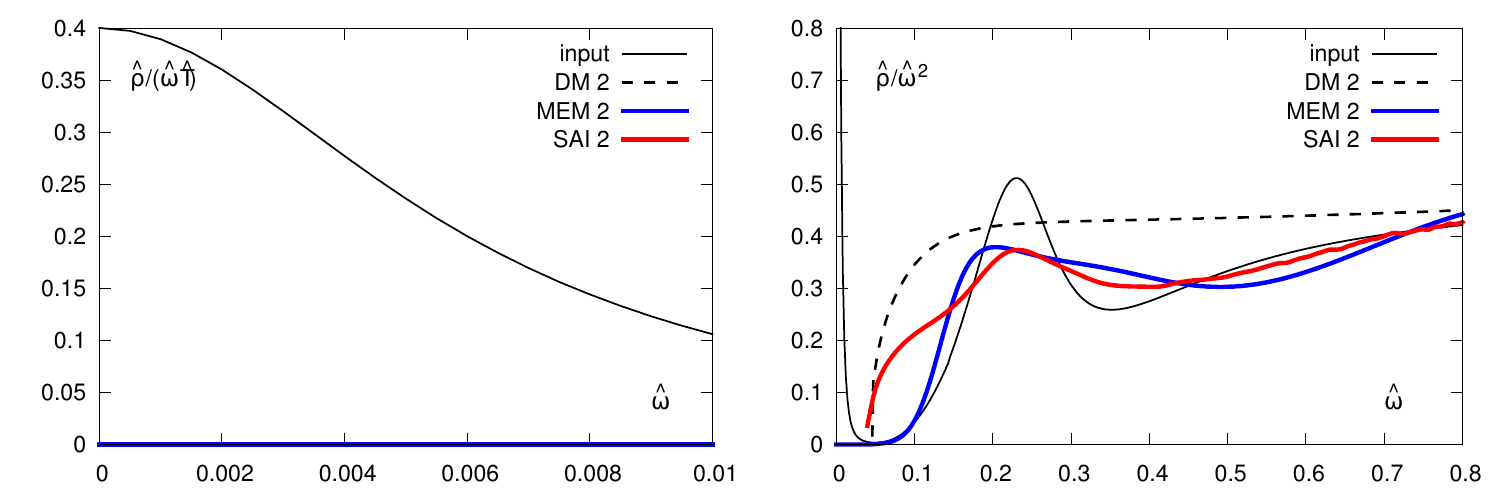} }
\caption{Dependences on default models  at $T>T_c$ with $N_{\tau}=48$ and $\epsilon=5\times10^{-3}$. \textit{Top two figures:} Results obtained using  $DM1$. \textit{Bottom two figures:} Results obtained using $DM2$. Both $DM1$ and $DM2$ are rescaled free Wilson spectral functions. The value of the quark mass $m$ is set to be 0.06 in $DM1$, and it is 0.02 in $DM2$.}
\label{dm12}
\end{figure}

\begin{figure}[!htpb]
\centerline{\includegraphics[width=0.8\textwidth]{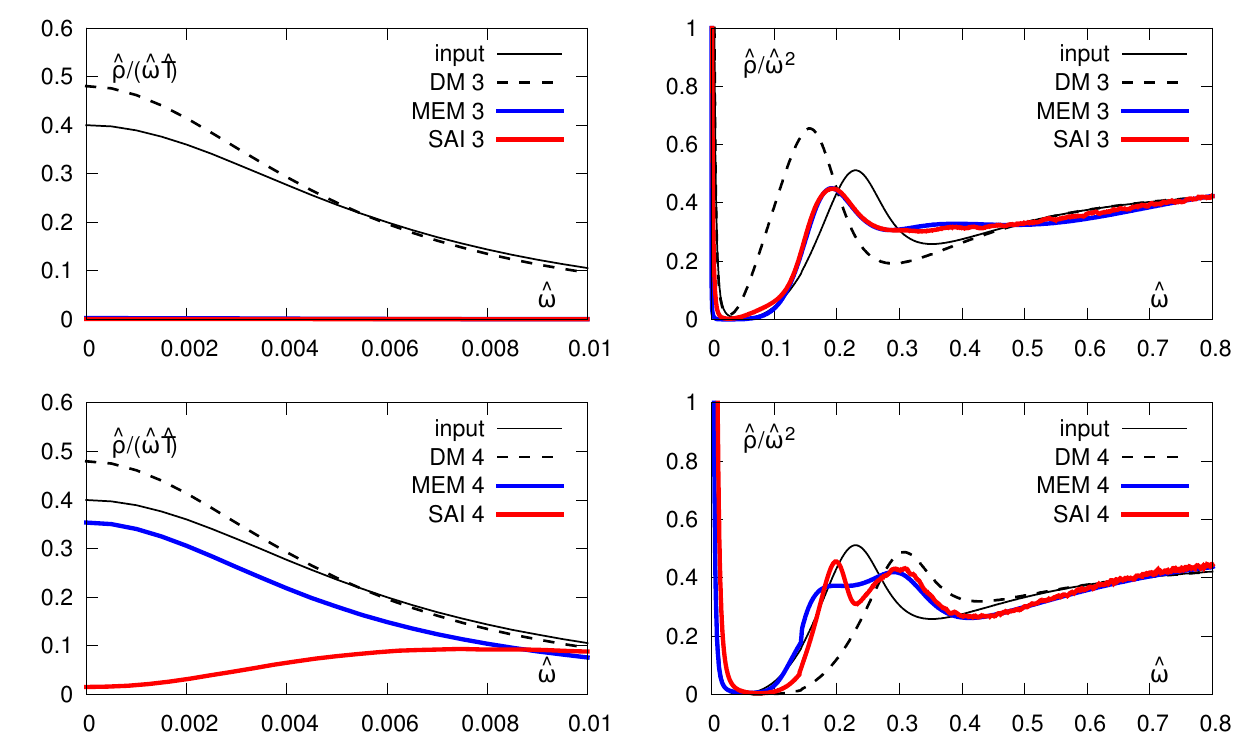} }
\caption{Dependences on default models at $T>T_c$ ($N_{\tau}=48$) with a noise level $\epsilon=5\times10^{-3}$. \textit{Top two figures:} Results obtained using $DM3$. \textit{Bottom two figures:} Results obtained using $DM4$. $DM3$ and $DM4$ have almost the same transport peak and large $\hat \omega$ part but have different resonance peak locations. In $DM3$ the resonance peak locates at $M=0.155$ while in $DM4$ it locates at $M=0.300$.}
\label{dm34}
\end{figure}

\begin{figure}[!htpb]
\centerline{\includegraphics[width=0.8\textwidth]{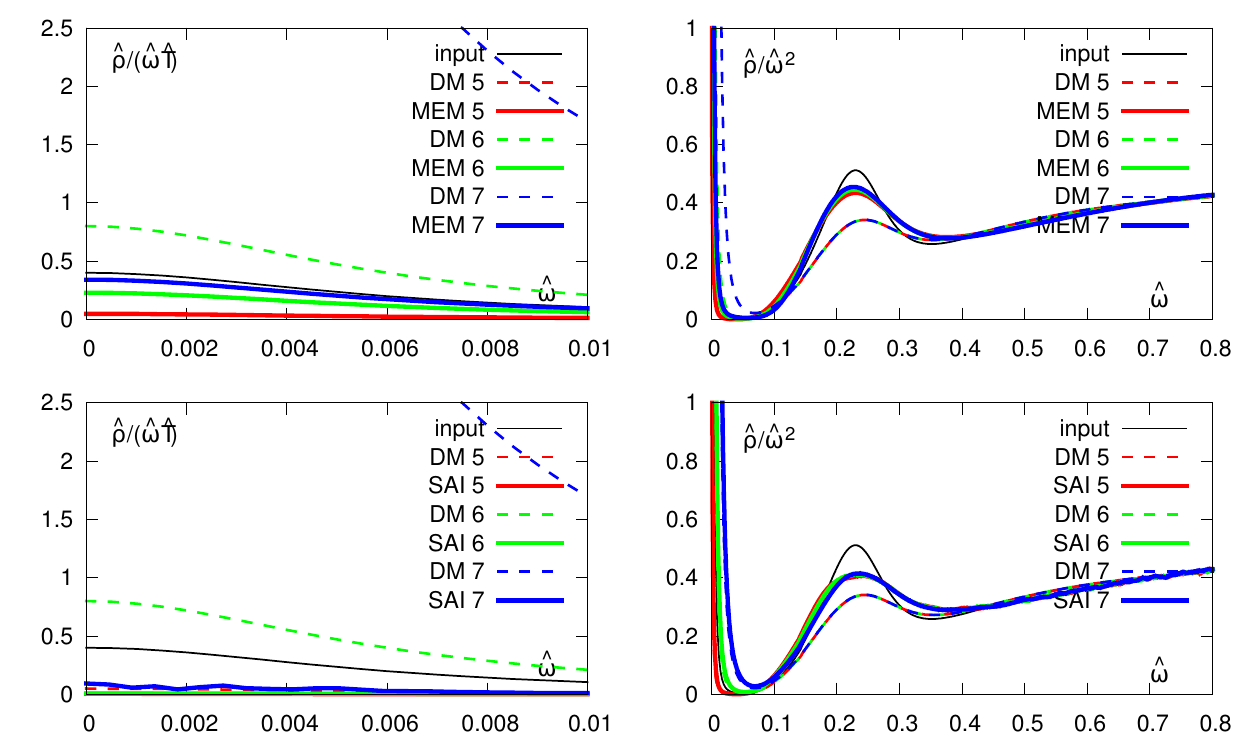} }
\caption{Dependence on default models tests at $T>T_c$ ($N_{\tau}=48$) in noise level $\epsilon=5\times10^{-3}$. \textit{Top two figures:} Results of MEM. \textit{Bottom two figures:} Results of SAI. $DM4-DM6$ have the same resonance peak and a large $\hat \omega$ part. For a transport peak they have the same width but a different height. $DM5$ has $c_{trans}=5\times10^{-5}/8$. $DM6$ has $c_{trans}=5\times10^{-5}\times2$. $DM7$ has $c_{trans}=5\times10^{-5}\times16$. Here $5\times10^{-5}$ is the $c_{trans}$ of the input spectral function.}
\label{dm567}
\end{figure}

\begin{figure}[htpb]
\centerline{\includegraphics[width=0.8\textwidth]{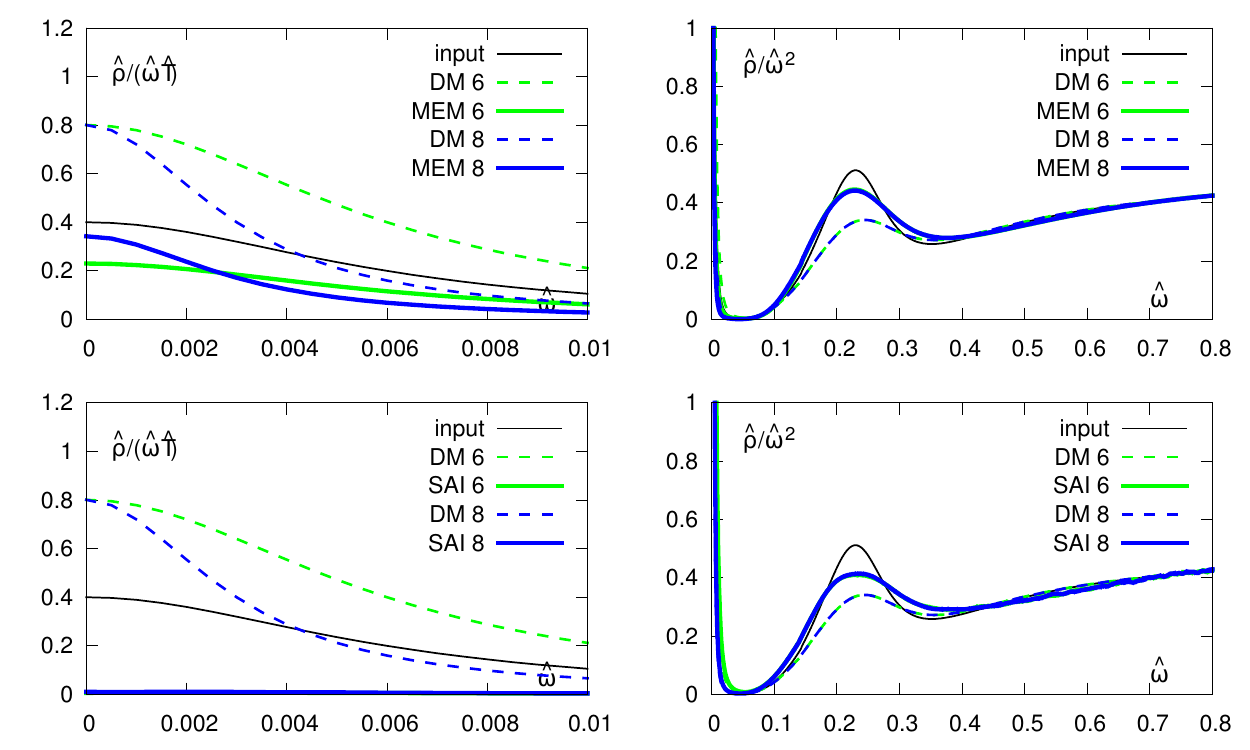} }
\caption{Dependence on default models tests at $T>T_c$ ($N_{\tau}=48$) in noise level $\epsilon=5\times10^{-3}$. \textit{Top two figures:} Results of MEM. \textit{Bottom two figures:} Results of SAI. $DM6$ and $DM8$ have the same resonance peak and large $\hat \omega$ part. For the transport peak they have the same height but a different width. $DM6$ has $\eta=0.006$. $DM8$ has $\eta=0.003$. }
\label{dm68}
\end{figure}

 \begin{table}[htb]
\begin{center}
\begin{tabular}{|c|c|c|}
		\hline
		Default model \ & Type \ & Parameters\\		\hline
		DM1 \ & $\hat{\rho}_{Wilson}$ \ & m=0.06 \\ 		\hline
                DM2 \ & $\hat{\rho}_{Wilson}$ \ & m=0.02 \\ 		\hline
		DM3 \ & $\hat{\rho}_{above}$ \ & $\eta=0.005,\hat{\omega}_2=\hat{\omega}_3=M=0.155$ \\ 		\hline
		DM4 \ & $\hat{\rho}_{above}$ \ & $\eta=0.005,\hat{\omega}_2=\hat{\omega}_3=M=0.300$ \\ 		\hline
		DM5 \ & $\hat{\rho}_{above}$ \ & $c_{trans}=5\times10^{-5}/8,\Gamma=0.25$ \\ 		\hline
		DM6 \ & $\hat{\rho}_{above}$ \ & $c_{trans}=5\times10^{-5}\times2,\Gamma=0.25$ \\ 		\hline
		DM7 \ & $\hat{\rho}_{above}$ \ & $c_{trans}=5\times10^{-5}\times16,\Gamma=0.25$ \\  		\hline
		DM8 \ & $\hat{\rho}_{above}$ \ & $\eta=0.003,\Gamma=0.25$ \\ \hline
	\end{tabular}
\end{center}
\caption{Parameters of the default models at temperature above $T_c$.}
\label{DM_Parameters}
\end{table}
  The parameters of these default models are listed in Table~\ref{DM_Parameters} and the results are summarized in Fig.\ref{dm12}-\ref{dm68}. The left panels of the figures show the transport peak in the small energy region while the right ones show spectral functions in larger energy region. 

%\clearpage{}

First let us see what happens when the default model is simply a rescaled free Wilson spectral function. From the left panels of Fig.~\ref{dm12} we see that the transport peaks given by both MEM and SAI are comparable to zero. This is due to the fact that the default models in this low frequency region are set to zero. While in the large energy region, as seen from the top-right panel of Fig.~\ref{dm12}, the peak locations of the resonance peaks given by both MEM and SAI differ a lot from that of the input spectral function and the rapidly rising part of the output spectral function at $\hat{\omega}$ around 0.1 just follows the behavior of $DM1$. This might arise from the issue that
$DM1$ does not cover a sufficiently small energy region, i.e. $\hat{\omega}\lesssim 0.1$. We then tried with $DM2$ which is same as $DM1$ but starts to be nonzero at a smaller threshold.  The results are shown in the bottom panel of Fig.~\ref{dm12}.  We can see that MEM still cannot reconstruct the peak location while SAI can give a peak-like structure which has a correct peak location, although the shape of resonance peak is not obvious. From this test one can learn that the default model should cover as wide a range as possible; otherwise, the missing part would have a fatal influence on the output spectral functions. In the following we will try to add an additional transport peak in the default model to see the effects.

We further check the default model dependencies using $DM3$ and $DM4$ composing of a transport peak, a resonance peak and a free Wilson spectral function. Here $DM3$ has a resonance peak location smaller while $DM4$'s peak location is larger than the input one. This is to say that we fix the transport peaks of these two default models to be similar to that of the input spectral function and vary the peak locations of the resonance peak in the default models. As seen from the top-right panel of Fig.~\ref{dm34} MEM and SAI give consistent output peak locations, i.e. $\hat \omega=$0.1915 for MEM and $\hat \omega=$0.1934 for SAI. And both the reconstructed peak locations are smaller than the input one, i.e. $M_2=$0.225. It can also be observed that the output peak locations move to a large energy region compared to that of the $DM3$, i.e. $M=$0.155. When using a default model that has a resonance peak location larger than the input spectral function, as shown in the bottom right panel of Fig.~\ref{dm34}, both MEM and SAI start to produce two separated peak/bump structures at $\hat{\omega}>0.1$, where locations of the first and second peaks/bumps are smaller than and close to that of the resonance peak in the $DM4$, respectively. The left panels of Fig.~\ref{dm34} show the transport peaks obtained from MEM and SAI analyses. From the top-left panel it can be seen that the transport peaks obtained from both MEM and SAI are compatible with zero while seen from the bottom-left panel MEM almost reproduces the transport peak while SAI still gives a much smaller intercept at a vanishing frequency. We thus conclude that the output spectral function extracted from correlates with $N_{\tau}=48$ and $\epsilon=5\times10^{-3}$ has a strong dependence on the peak location of the resonance peak in the $DM$. And the reconstruction of the resonance part also has considerable influence on the reconstruction of the transport peak. However, the tendency of the resonance peak location in the output spectral function indicates that the real resonance peak is located in between the peaks in the $DM3$ and $DM4$. We will then try default models with the resonance peak location lying in between that of $DM3$ and $DM4$ as follows.

%
%
%To obtain correct peak location of the resonance peak, one should try default models at different peak-locations and see when the peak-location in the output SPF is same as the one in the default model. The peak-location obtained at that point should be a good estimation of the true value.

In Fig.~\ref{dm567} we show the MEM and SAI results obtained using $DM5$, $DM6$ and $DM7$ which have the same resonance peak locations and different heights of the transport peak at $\hat{\omega}=0$ as the input spectral function.  From the right panels of Fig.~\ref{dm567} we see that the continuum part and resonance peak are reconstructed very well by both SAI and MEM. In the right top panel the resonance peak locations given by MEM using $DM5$, $DM6$ and $DM7$ are $\hat \omega=$0.232, 0.229 and 0.226, respectively while in the bottom-right panel SAI gives $\hat \omega=$0.243, 0.233 and 0.235, respectively. All are close to the input value of the peak location $M_2=$0.225. However, the default model dependence on the transport peak is still quite strong as seen from the left panels. From the MEM analyses (top-left panel) one is able to see that the height of the output transport peak approaches the input one, while from the SAI analyses (bottom-left panel) the output transport peak grows slightly as the default model but is still comparable with zero.

The analyses done in Fig.\ref{dm34} and Fig.~\ref{dm567} suggest that the resonance peak location is reproduced correctly when 
resonance peak locations of the default model and the output spectral function are comparable. And the upper bound for the height of the output transport peak (with the correct width of the transport peak in the DM) obtained from the MEM analysis can be a good estimate of its real value. However, it has to be noted that the width $\eta$ of the transport peak in the default model used in these tests is fixed to be the same as the input. We will check the dependence on $\eta$ in the following tests.

In Fig.~\ref{dm68} $DM8$ is same as $DM6$ except for the width of the transport peak. The right panels show that in both the MEM and SAI analyses the resonance part is reproduced well again as above and the variation of the transport peak in the default model has a mild influence on the reconstruction of the resonance peak. The reconstruction of the transport peak is shown in the left panel of Fig.~\ref{dm68}. Here the width of the transport peak in $DM6$ and $DM8$ is $\eta=0.006$ and $\eta=0.003$, respectively. The output width given by MEM shown in the top-left panel is $\eta=0.00612$ using $DM6$ and $\eta=0.00302$ using $DM8$ (obtained by least-$\chi^2$ fitting in the small $\hat{\omega}$ range using a Lorentz peak as the ansatz). In the bottom-left panel SAI fails to reconstruct a transport peak for these two default models. We can see that MEM just repeats the width of the transport peak in the default model and only when the width is known, one is able to reproduce the right transport peak right from MEM.

\section{Analysis with Lattice QCD Data}
\label{real_data_analysis}

In this section we will present the charmonia spectral functions in the pseudoscalar ($\eta_c$) and vector ($J/\psi$) channels extracted using the SOM, SAI and MEM. The correlators used in our analyses are taken from Ref.\cite{Ding:2012sp} and here we only focus on the correlators computed on the finest lattices, i.e.  $128^3\times96$ and $128^3\times48$ corresponding to temperatures at $0.75T_c$ and $1.5T_c$. In our analyses we constrain the frequency range $\hat \omega_{max}=4$, or $\omega_{max}=75.88$ GeV (lattice spacing $a^{-1}=18.97$ GeV). As the correlators calculated on lattices suffer from the lattice cutoff effects, which would manifest themselves at small distances or large energy range, we thus would abandon a first few points of the correlates in the short distance and set the reference imaginary time $\hat \tau_0=4$  in our analyses. 

\subsection{Spectral functions for the pseudoscalar channel}

Firstly, we consider the pseudoscalar channel at a temperature $T<T_c$. In this case we use four different DMs. These four default models are the same with the ones used in Sec.~\ref{Below_$T_c$}. The only difference is that the free continuum in the large $\omega$ part is replaced with a free Wilson spectral function in the pseudoscalar channel with the quark mass $m=0.06$ in lattice unit [see Eq.(\ref{wilson_peak})].
\begin{figure}[htb]
\centerline{\includegraphics[width=0.8\textwidth]{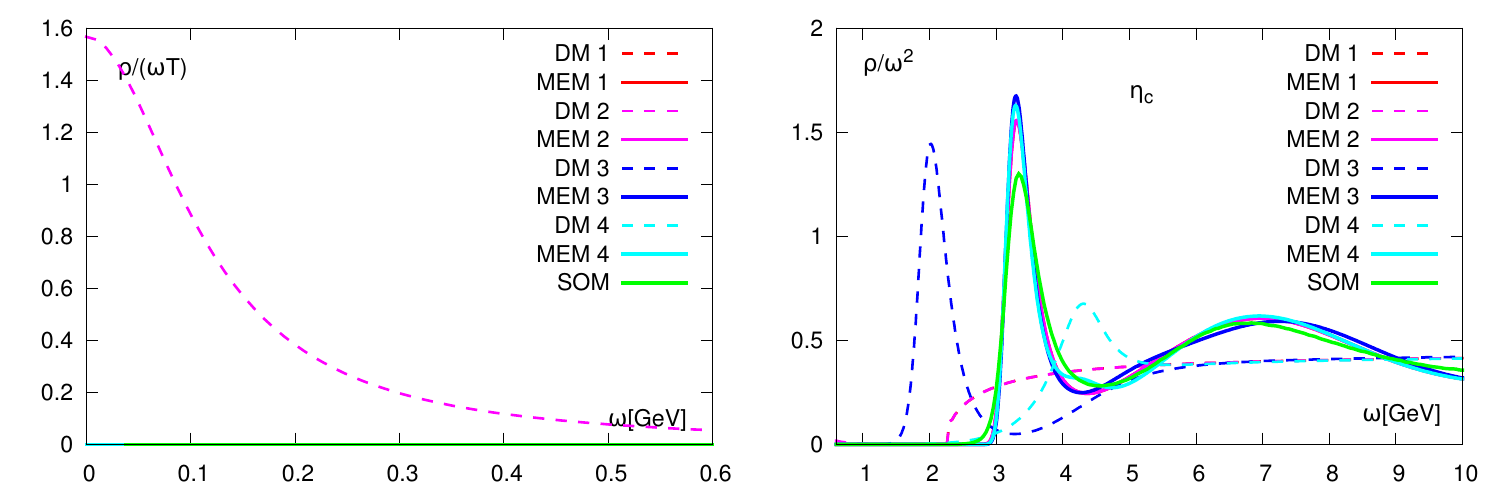} }
\centerline{\includegraphics[width=0.8\textwidth]{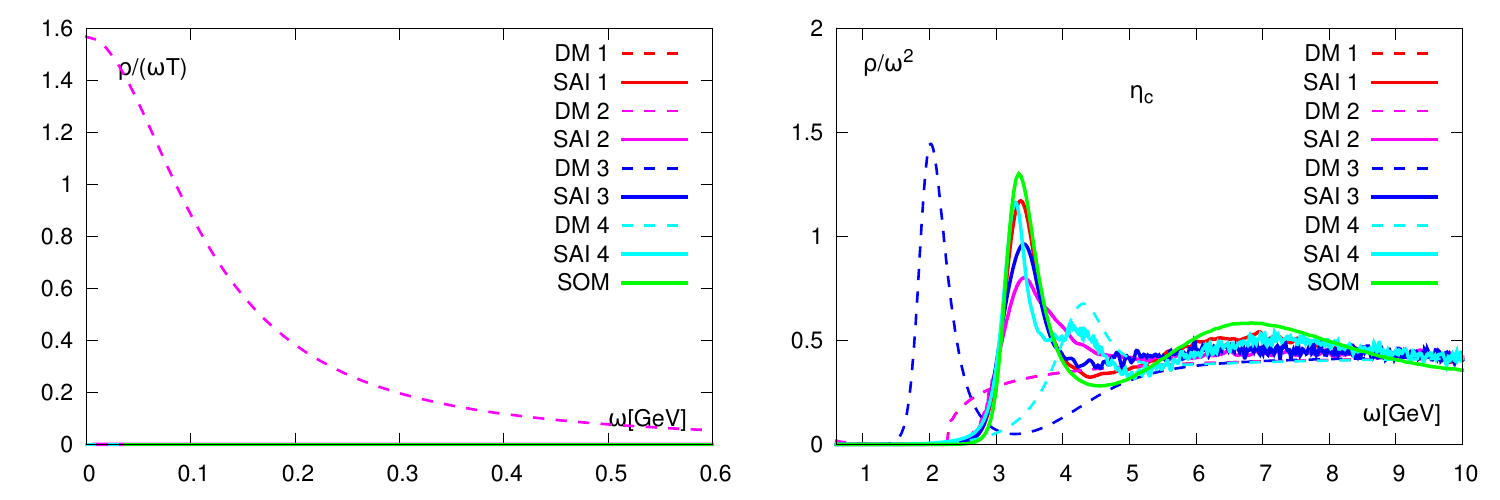}}
\caption{Spectral functions obtained by SOM, SAI and MEM at $0.75T_c$ for the pseudoscalar channel ($\eta_c$). \textit{Top two figures}: The results of MEM and SOM. \textit{Bottom two figures}: The results of SAI and SOM.}
\label{PS_075Tc}
\end{figure}

The results are shown in Fig.~\ref{PS_075Tc}. It can be observed from the right panels of Fig.~\ref{PS_075Tc} that the first resonance peaks obtained using both MEM and SOM are default model independent and are very stable. The peak locations obtained by MEM with $DM1$-$DM4$ are at $\omega=3.310, 3.310, 3.300, 3.291$ GeV, respectively. Since MEM with $DM2$ gives almost the same result as with $DM1$  one cannot distinguish in the figure. 
On the other hand SAI gives a resonance peak location at $\omega=3.366, 3.418, 3.416, 3.289$ GeV, respectively for $DM1$-$DM4$. Although there are no default models used in the SOM analysis the resonance peak location obtained from the SOM is at $\omega=3.377$ GeV which is compatible with those obtained from the MEM and SAI.
For the transport peak shown in the left panels of Fig.~\ref{PS_075Tc} we can see that the transport contributions in the output spectral function from all these three methods are compatible with zero even a default model ($DM1$) with a nonzero transport peak is used.

\begin{figure}[htb]
\centerline{\includegraphics[width=0.8\textwidth]{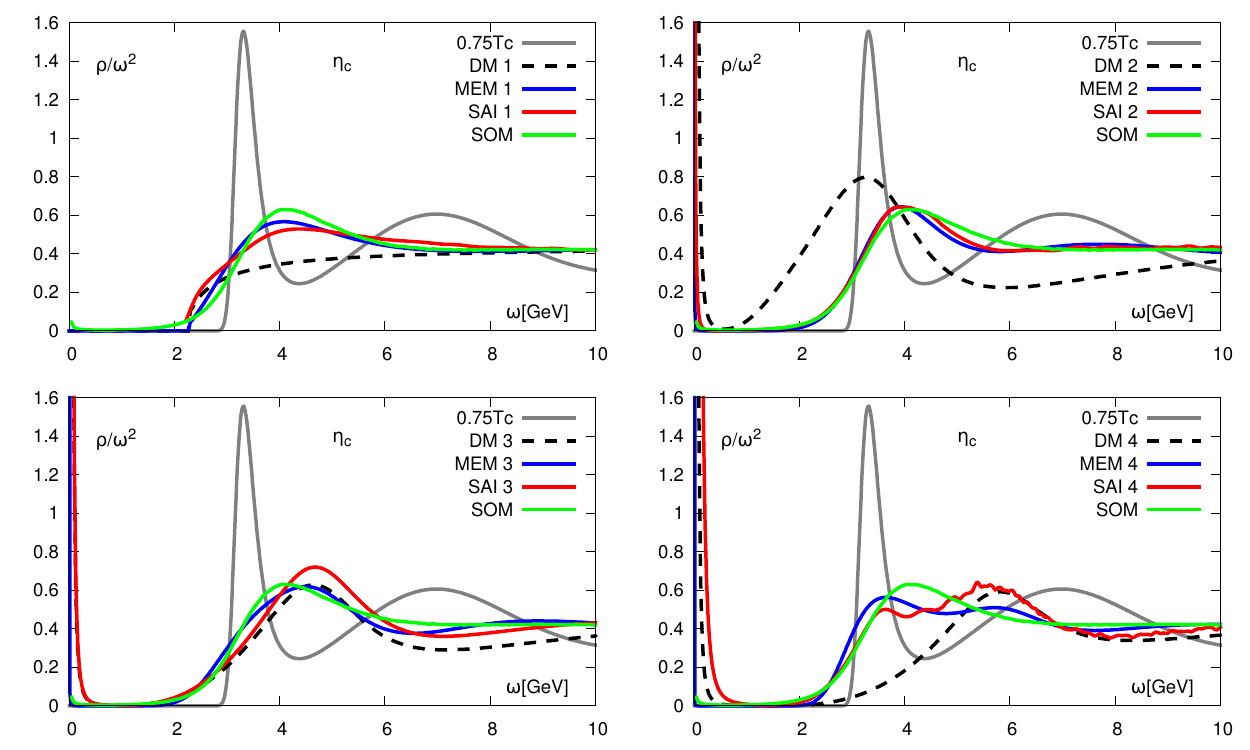} }
\caption{Dependence of the location of the possible $\eta_c$ resonance peak on various DMs at $1.5T_c$. }
\label{PS_M_dep}
\end{figure}

\begin{figure}[htb]
\centerline{\includegraphics[width=0.8\textwidth]{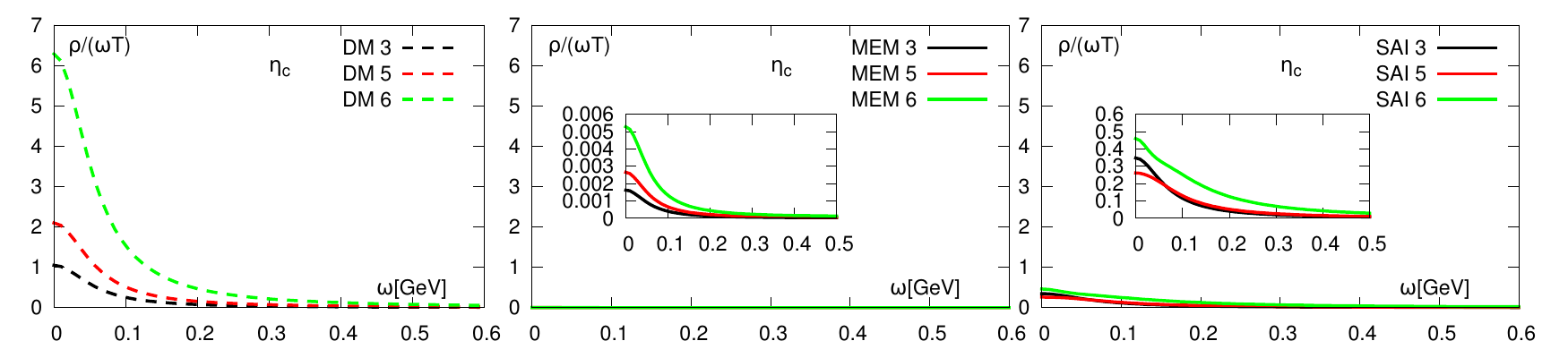} }
\caption{Dependence of the transport peak on the DMs at $1.50T_c$ in the pseudoscalar ($\eta_c$) channel.}
\label{PS_150Tc_trans}
\end{figure}

Then we move on to the analyses of spectral function in the pseudoscalar channel at $1.50T_c$, and the results are shown in Fig.~\ref{PS_M_dep} and Fig.~\ref{PS_150Tc_trans}. Here the $DM1$ we use is simply a rescaled free Wilson spectral function. The other DMs are of a general type: $DM(\omega)=\hat \rho_{above}(M,\Gamma,\omega,...)$ but with different parameters, which have three parts: one transport peak, one resonance peak and one free Wilson spectral function. $DM2$, $DM3$ and $DM4$ have the same transport peak and large $\omega$ parts but different resonance parts. To construct the transport peak we fix $\eta=0.003, c_{trans}=\pi\times10^{-5}$. The resonance peaks are located at $\omega=3.2249, 4.5528, 5.6910$ GeV for $DM2$, $DM3$ and $DM4$, respectively. The peak location of the resonance peak in $DM2$ is chosen to be close to the results obtained at $0.75T_c$. For comparison we also show the spectral function obtained by MEM with DM1 at $0.75T_c$ denoted as ``0.75 $T_c$" in Fig.~\ref{PS_M_dep}.

%We summarize the results in Fig.\ref{PS_M_dep} and Fig.\ref{PS_150Tc_trans}. 

Firstly we show the reconstructed spectral function in the large energy region in Fig.~\ref{PS_M_dep}.
It can be seen that SOM gives a single resonance peak located at $\omega=4.211$ GeV. From the top-left panel we see that with a rescaled free Wilson spectral function as the default model both MEM and SAI do not produce a suddenly rising resonance peak as shown in Fig.~\ref{dm12}, and they even reconstruct a peak structure already with this simple default model. This indicates that the free Wilson spectral function has covered at least most of the energy range that the real spectral function covers in the energy region relevant for the resonance peaks. The results of MEM and SAI based on $DM2$-$DM4$ show that as the resonant peak location in the default models increases it also increases in the outputs and finally splits into two resonance peaks in the bottom-right panel.

Going by Fig.\ref{PS_M_dep}, for all the methods, at best we can describe the broad and low resonancelike peak structure in the $\eta_c$ channel at $\omega=4.553$ GeV, which is about 40\% larger than observed at $0.75T_c$. This suggests that in a gluon plasma at $1.5T_c$ $\eta_c$ does not exist as a clearly identifiable bound state.

Finally, we consider the spectral function in the very small energy region. We fix the resonance part the same as $DM3$ in the following and only change the default model in the very small energy region. We choose the width of the transport peak to $\eta=0.003$ and vary the height factor $c_{trans}=\{\pi\times10^{-5}, 2\pi\times10^{-5}, 6\pi\times10^{-5}\}$. The results are shown in Fig.\ref{PS_150Tc_trans}. First let us look at the MEM results shown in the middle. We can see from that when the peak height of the transport peak increases, the output one in MEM analysis also increases. However, the output values are quite small compared with the DMs shown on the left. Furthermore, the increasing trend in the output is not so fast as the DMs. This strongly indicates that there does not exist a transport peak, which is expected in the pseudoscalar channel~\cite{Karsch:2003wy,Aarts:2005hg,Umeda:2007hy}. In fact, the transport peaks obtained here are within the error of the correlators. To see this we calculate the contribution to the correlator at the middle point $\tau T=0.5$ from the largest transport peak obtained by MEM using $DM6$. Integrating over the region $\omega \in [0,0.598]$ GeV, we obtain the contribution $G_{trans}(\tau T=0.5)=3.43\times 10^{-9}$, which is smaller than the error at the middle point $\delta G(\tau T=0.5)=7.91\times 10^{-8}$. While for the SAI results shown in the right panel of Fig.~\ref{PS_150Tc_trans}, we can see that the intercept at $\omega=0$ is almost 100 times larger than MEM results. However, the intercept is still very small compared to that of the default model.  Due to the stochastic nature of the SAI it is most likely that the spectral function in the small energy region is compatible with zero as seen in the MEM outputs.

%\begin{figure}[htb]
%\centerline{\includegraphics[width=0.5\textwidth]{error_ps48_small.pdf} \includegraphics[width=0.5\textwidth]{error_ps48_large.pdf}}
%\caption{The error analysis of the $\eta_c$ spectral function in SOM, SAI and MEM at $1.5T_c$. \textit{Left}: The transport peak. \textit{Right}: The resonance peak.}
%\label{error_analysis_ps48}
%\end{figure}

\subsection{Spectral functions for the vector channel}

\begin{figure}[htb]
\centerline{\includegraphics[width=0.8\textwidth]{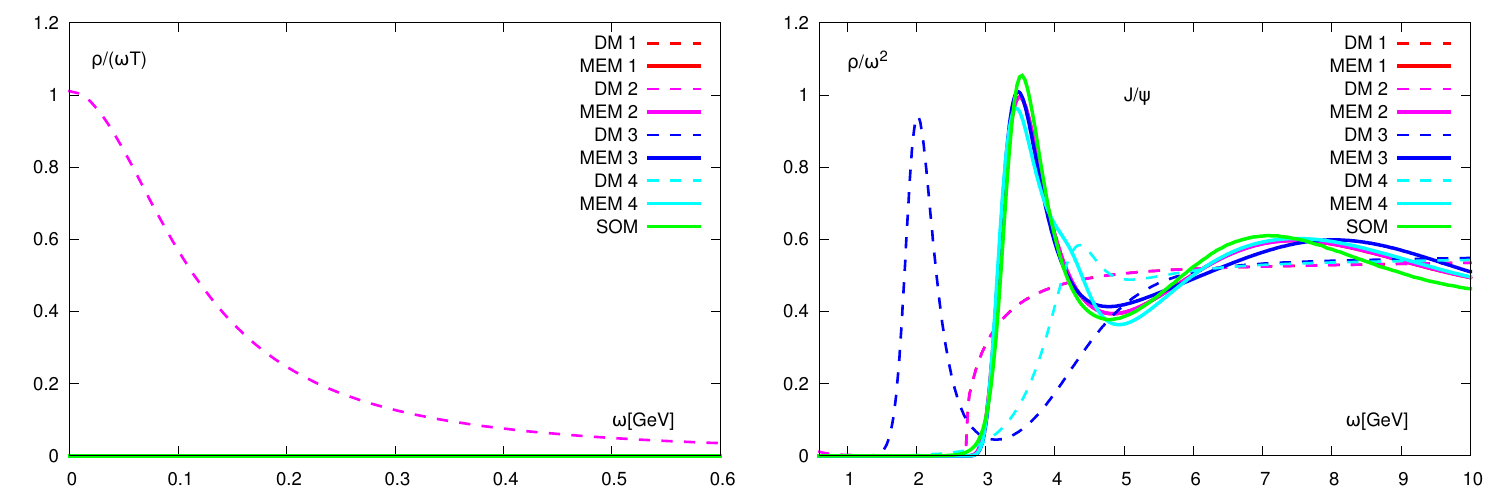} }
\centerline{\includegraphics[width=0.8\textwidth]{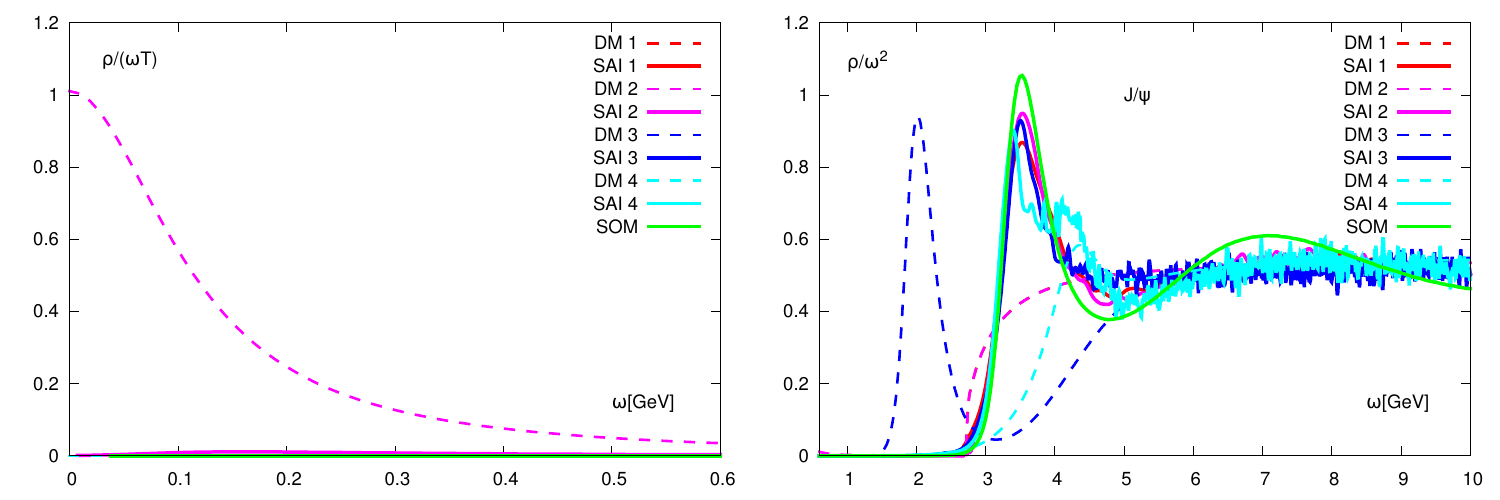} }
\caption{Spectral functions obtained by SOM, SAI and MEM at $0.75T_c$ for the vector channel ($J/\psi$). \textit{Top two figures}: The results of MEM and SOM. \textit{Bottom two figures}: The results of SAI and SOM.}
\label{VV_075Tc}
\end{figure}

\begin{figure}[htbp!]
\centerline{\includegraphics[width=0.8\textwidth]{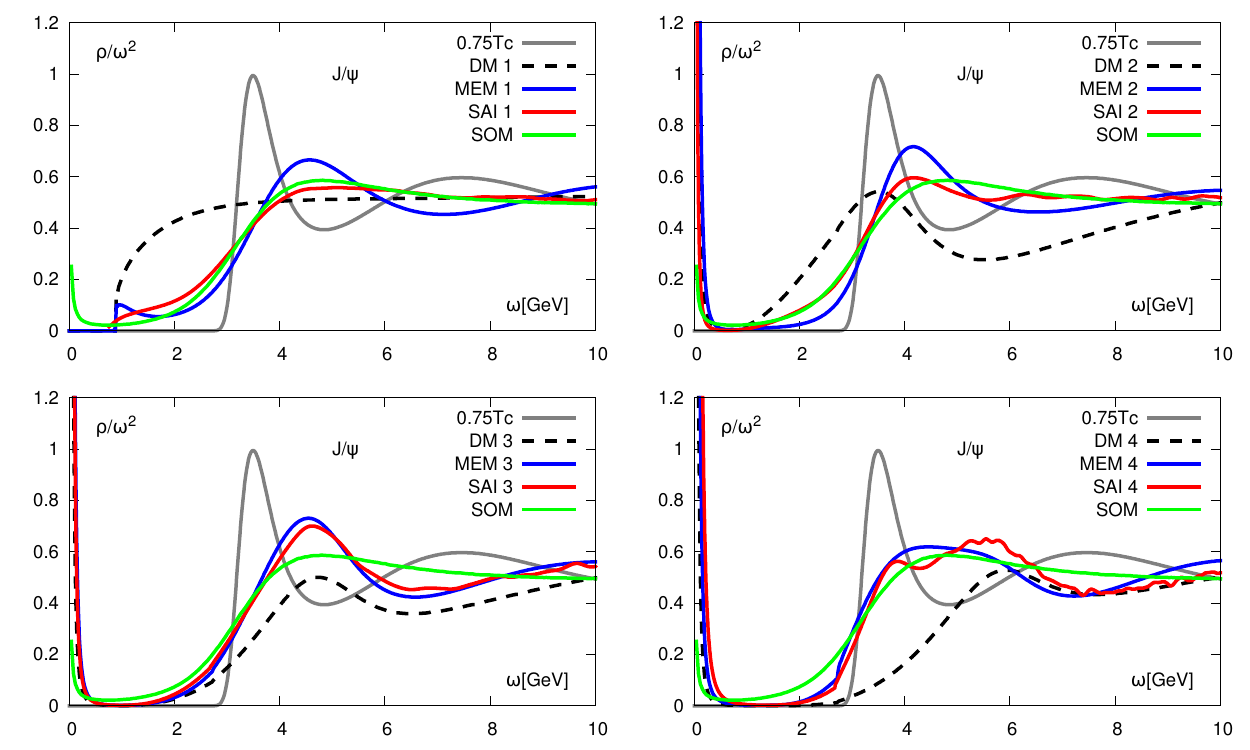} }
\caption{Dependence of the location of the possible $J/\psi$ resonance peak on various DMs at $1.50T_c$.}
\label{VV_M_dep}
\end{figure}

\begin{figure}[htbp!]
\centerline{\includegraphics[width=0.8\textwidth]{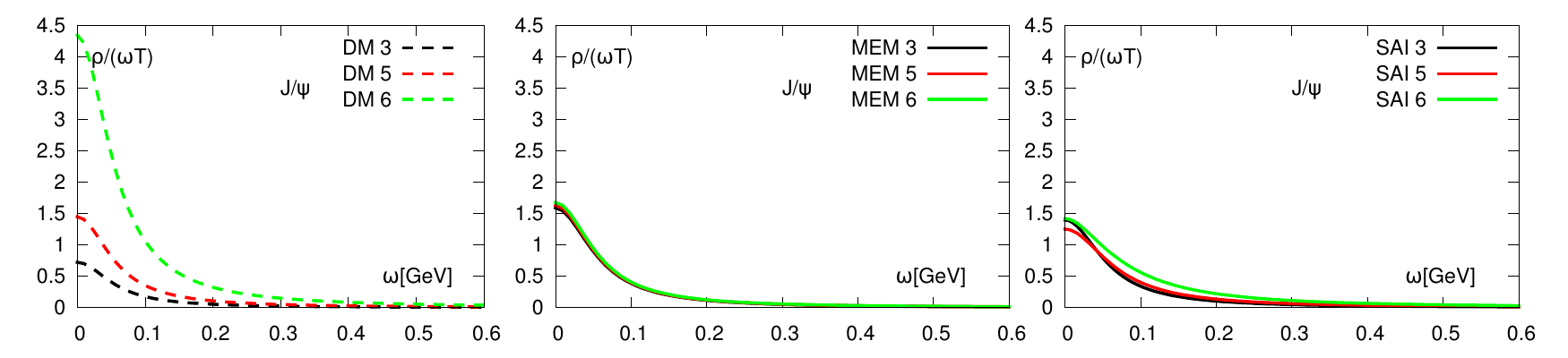} }
\caption{The dependence on the transport peak in the DMs at $1.50T_c$ for the vector channel.}
\label{VV_150Tc_trans}
\end{figure}

In this section we show the results of the spectral functions for the vector channel. The results at $T<T_c$ are shown in Fig.~\ref{VV_075Tc}.
 At $T<T_c$ the default models used here are the same as those used in the pseudoscalar channel except that the free Wilson spectral function is updated to the one in the vector channel. We can see that as in the pseudoscalar channel there does not exist any transport peak in the vector channel at $T<T_c$. As for the resonance part, MEM gives peak locations at $\omega=3.490, 3.490, 3.472, 3.443$ GeV obtained using  $DM1$, $DM2$, $DM3$  and $DM4$, respectively, as shown in the top panel of Fig.~\ref{VV_075Tc}. In the bottom panel SAI also shows a stable resonance peak location at $\omega=3.529, 3.529, 3.503$ GeV from $DM1$, $DM2$ and $DM3$. It is worthy to mention that the resonance peak location from the SOM is at $\omega=3.528$ GeV and it is quite compatible with the results obtained from MEM and SAI.

The results of spectral function at $T=1.5~T_c$ are presented in Fig.~\ref{VV_M_dep} and Fig.~\ref{VV_150Tc_trans}. Here the default models used are the same as those in the pseudoscalar channel except for two modifications. Firstly, we replace the rescaled free Wilson spectral function by the one in the vector channel, and secondly, the smallest resonance peak location $\omega=3.2249$ GeV is replaced with $\omega=3.4146$ GeV. This is because the resonant peak location at the below $T_c$ temperature obtained in the vector channel is larger. From the analyses shown in Fig.\ref{VV_M_dep}, for all the methods, at best we can obtain the broad and low resonancelike peak structure in the $J/\psi$ channel at $\omega=4.553$ GeV, which is about 30\% larger than observed at $0.75T_c$. This suggests that in a gluon plasma at $1.5T_c$ $J/\psi$ does not exist as a clearly identifiable bound state.

Finally we consider the transport peak. As seen from the middle panel of Fig.~\ref{VV_150Tc_trans} a very stable transport peak exists with a intercept of $\frac{\rho}{\omega T}\sim 1.6$ at $\omega=0$. This corresponds to $2\pi TD\sim8$ which is about 3 times larger than that in Ref.~\cite{Ding:2012sp}(in Ref.~\cite{Ding:2012sp} the corresponding default model has $\eta\sim0.008$ while in this paper we use $\eta=0.003$). The increasing of the transport peak in the default model does not affect the one in the output much. This also holds in SAI, although the outputs are more sensitive to the DMs. We calculate the contribution of the transport peak to the correlator at the middle point using the output spectral function in the range $\omega \in [0,0.598]$ Gev based on $DM6$ in MEM and SAI. They are $1.02\times 10^{-06}$ and $ 1.16\times 10^{-06}$ respectively which is much larger than the error in the middle point $4.29\times 10^{-08}$. So we can believe that this transport peak obtained from MEM and SAI cannot be generated from the error of the correlators.

\subsection{Reliability of the existence of resonancelike peak structures}

In this section we examine the reliability of the existence of resonancelike peak structures at 0.75$T_c$ and 1.5$T_c$. 
Firstly, we study the significance of the strength of the resonancelike peak structure estimating errors based on Eq.(\ref{variance_average}). The motivation here is the following— in the ideal case of a delta functionlike resonance structure an error estimate based on Eq.(\ref{variance_average})  provides the error on amplitude of the delta functionlike resonance and helps us to judge its significance over the continuum part of the spectral function. Next, we test the systematics in the reconstruction of the spectral function at 1.5$T_c$ by comparing it with the spectral function extracted from the so-called reconstructed correlator at 0.75$T_c$, i.e. from the correlation function that consists of the spectral function at 0.75$T_c$ but convoluted with the integrand kernel corresponding to 1.5$T_c$.

\begin{figure}[htb!]
	\includegraphics[width=0.45\textwidth]{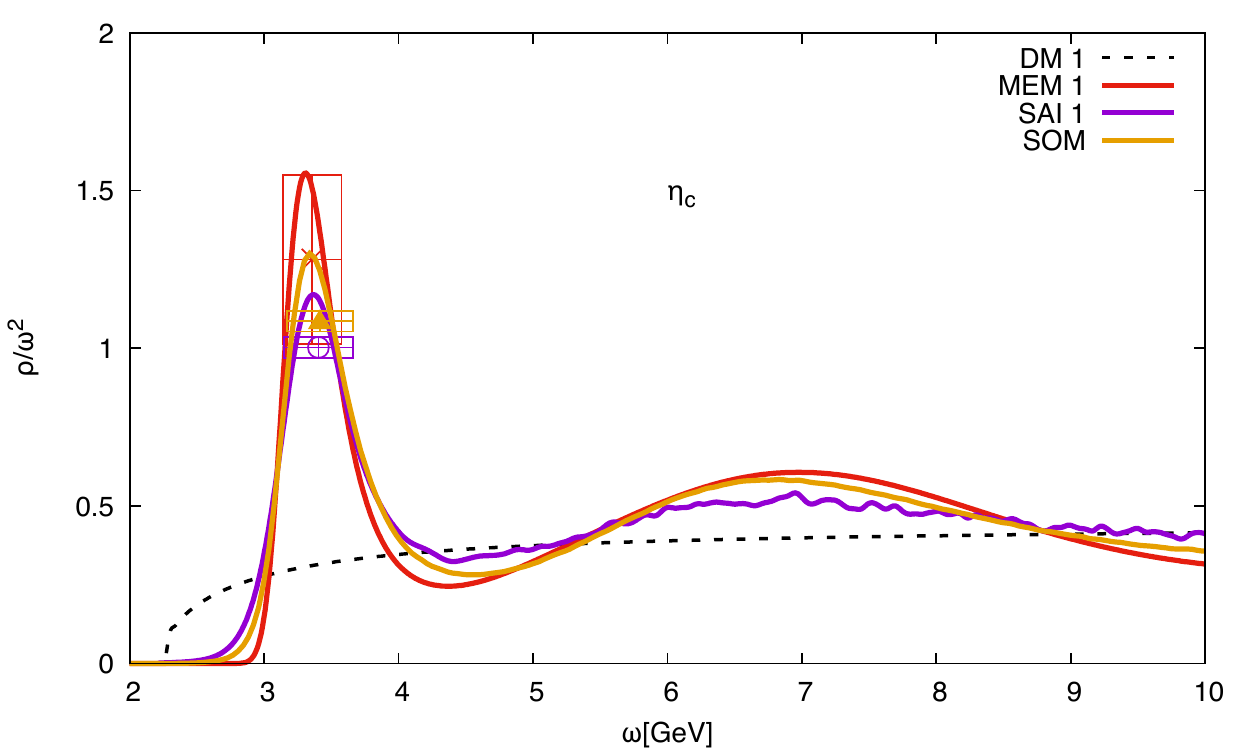}~\includegraphics[width=0.45\textwidth]{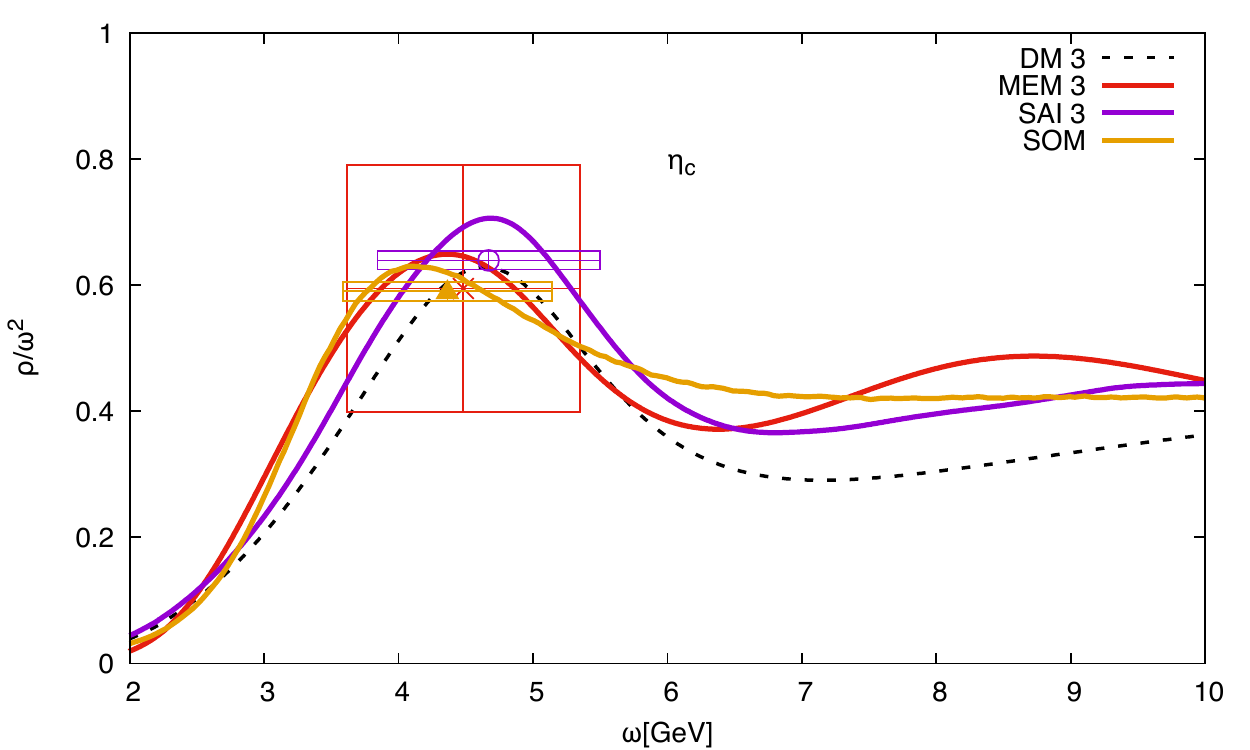}
	\caption{\textit{Left}: Significance of the strength of the resonancelike peak structure in $\eta_c$ spectral function obtained using SOM, SAI and MEM at $0.75T_c$. \textit{Right:} Same as left, but for the case at $1.5T_c$.}
	\label{error_analysis_ps}
\end{figure}

\begin{figure}[htb!]
\centerline{\includegraphics[width=0.45\textwidth]{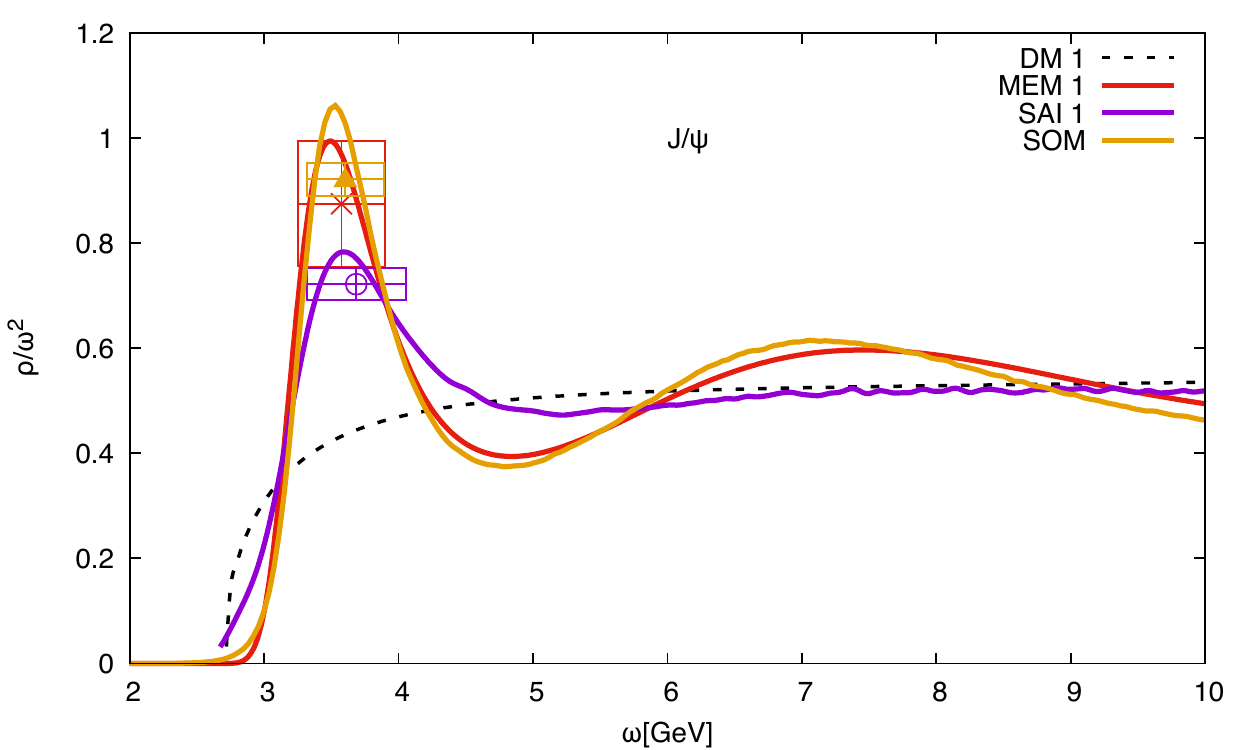} ~\includegraphics[width=0.45\textwidth]{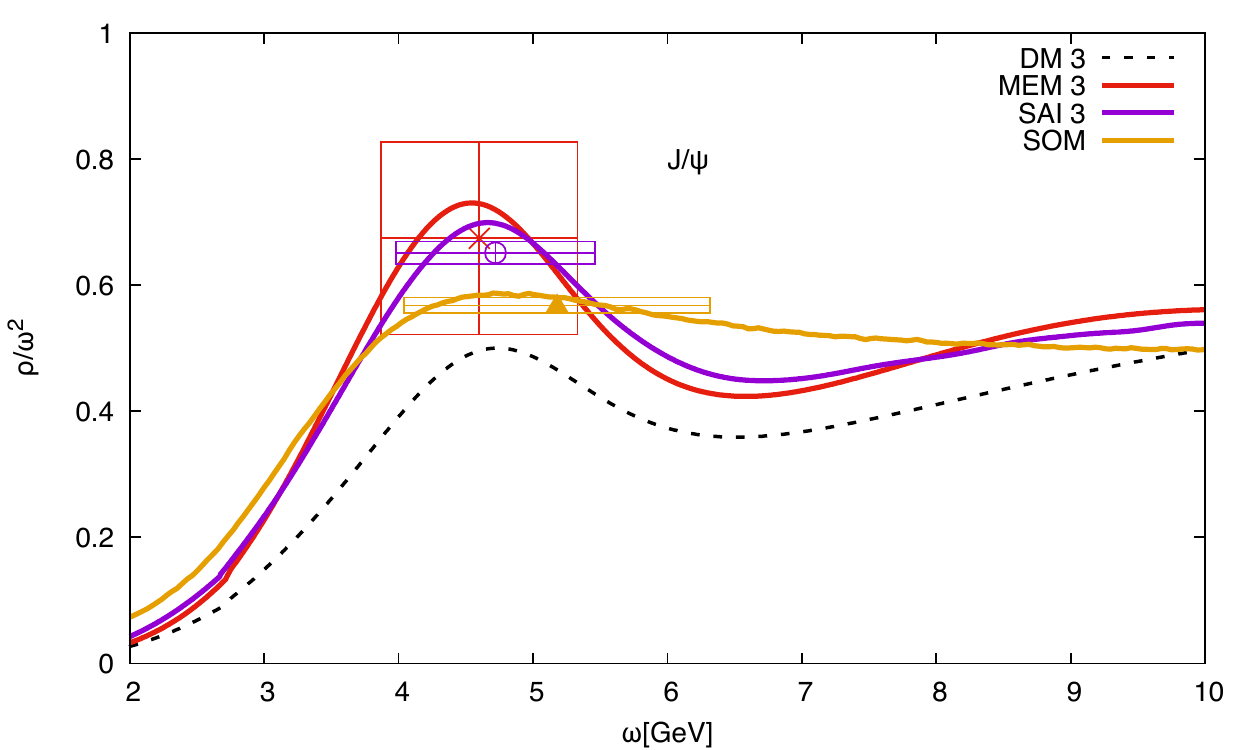}}
\caption{Same as Fig.~\ref{error_analysis_ps}, but for the vector channel.}
\label{error_analysis_vv}
\end{figure}

The rectangular boxes in Figs.~\ref{error_analysis_ps}  and ~\ref{error_analysis_vv} show our estimates for the significance of the existence of the resonancelike peak structures in the pseudoscalar and vector spectral functions, respectively. The width of the box characterizes the frequency interval $\mathit{I}$ over which SPF is averaged. The frequency-ranges, $I$, are chosen to be the full-widths at half maxima of the resonancelike peaks over the continuumlike structures, determined from the differences of the locations of maxima and the immediate minima to the right of the maxima. Along the y-axis the centers of the boxes are located at the mean values of the areas of the spectral functions integrated over frequency-ranges, $I$, and vertical half-extents of the boxes provide the one sigma-uncertainties on those integrated areas.  It can also be seen that  amplitudes of the $\eta_c$ and $J/\psi$  resonancelike structures are statistically significant at $0.75T_c$, but at $1.5T_c$ those statistical significances are questionable.   

From Figs.~\ref{error_analysis_ps} and ~\ref{error_analysis_vv}  we can see that the estimated uncertainties on the amplitude of the resonancelike peak are larger for MEM than that for SAI and SOM. To understand this we have further checked that even for a fixed value of $\alpha$, around its most probable value, the estimate for MEM gives a larger error than for SOM and SAI, and is not caused by the averaging over $P[\alpha|G,D]$. This leads us to speculate that the larger estimate of error for MEM might be due to its mean-field nature and the assumption of the sharp Gaussian approximation of  $P[\rho|G,D,\alpha]$ [$c.f.$ Eq.~(\ref{variance_alpha}] and ~(\ref{variance_alpha_mem})), whereas $P[\rho|G,D,\alpha]$ is sampled exactly for SOM and SAI.

\begin{figure}[h]
\centerline{\includegraphics[width=0.5\textwidth]{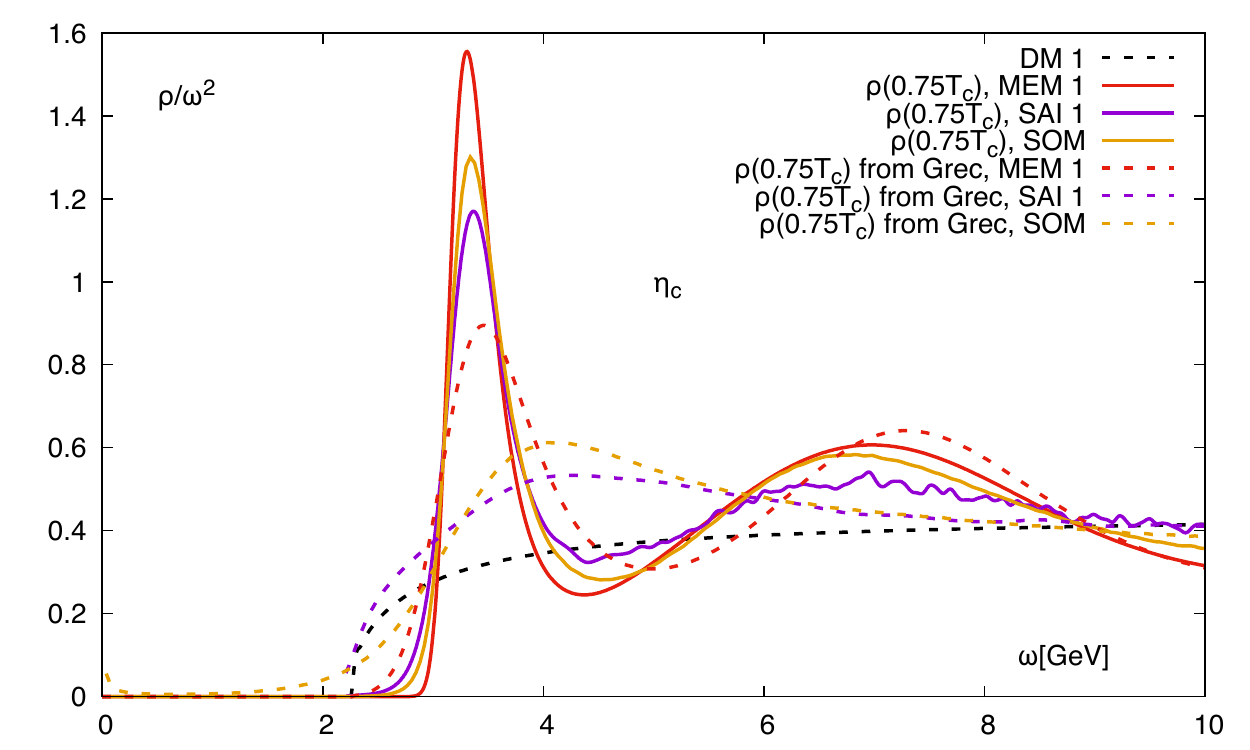} \includegraphics[width=0.5\textwidth]{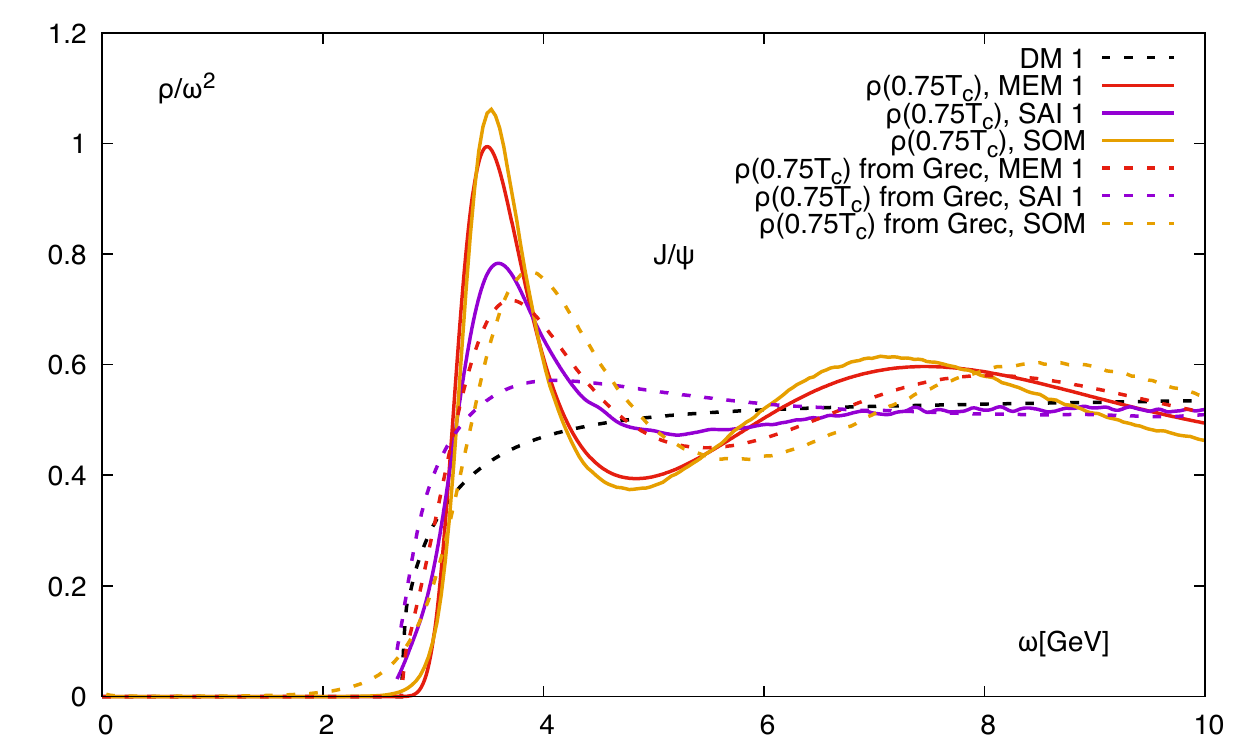}}
\caption{Spectral functions in the pseudoscalar channel (\textit{left}) and vector channel (\textit{right}) obtained by three different methods from the reconstructed correlators $G_{rec}(\tilde \tau,T=1.5T_c;T’=0.75T_c)$. The default models used here are the same as the first ones of those used in the analysis of $\eta_c$ and $J/\psi$ correlators at $0.75T_c$. The solid curves are results obtained by corresponding methods from correlators at $0.75T_c$ with $DM1$ for comparison.}
\label{recons_fig}
\end{figure}

Next, we consider reconstructed correlators~\cite{Ding:2012sp}
\begin{equation}
\label{recons_corr}
G_{rec}(\tilde \tau,T;T')=\sum_{\tilde \tau'=\tilde \tau,\Delta \tilde \tau'=N_{\tau}}^{N_{\tau}'-N_{\tau}+\tilde \tau}G(\tilde \tau',T'),
\end{equation}
where $\tilde \tau=\tau/a$, $\tilde \tau'=\tau'/a$, $T=1.5T_c$, $T'=0.75T_c$. We summarize the output spectral functions obtained from MEM, SOM and SAI in Fig.~\ref{recons_fig}.  We use the same default models as the first ones of those used in the analysis of $\eta_c$ and $J/\psi$ correlators at $0.75T_c$. For comparison we include the results obtained at $0.75T_c$ shown as solid curves. The results obtained from the reconstructed data are labeled with ``$\rho(0.75T_c)\ from\ G_{rec}$".  By comparing the results obtained from the correlators at 0.75$T_c$ with $N_\tau=96$ and from the reconstructed correlators with $N_\tau=48$ we can see that the results from MEM have smaller $N_\tau$ dependences while results from SOM and in particular SAI suffer from the reduction in $N_\tau$. It is also worth to mention that the differences between the output spectral functions from the original correlators and the reconstructed correlators are larger in the pseudoscalar channel than those in the vector channel. This is due to the fact that the noise-to-signal ratio of the correlators in the former case is around 80\% larger at the largest distance, and the insufficient quality of the data is also indicated by unphysical nonzero contributions at $\omega\sim 0$ in the pseudoscalar spectral function from $G_{rec}$ obtained using SOM~\footnote{In the above described analyses we used the covariance matrix of the reconstructed correlators which reflects the statistical uncertainties and correlations at $0.75T_c$. We also examined the role of the covariance matrix itself in all three methods. We test what happens when we use the full covariance matrix at $1.5T_c$ and rescale it by the ratio of the mean values of the reconstructed correlator at $0.75T_c$ to the original correlator at $1.5T_c$. These tests show that the explicit role of the covariance matrix for these spectral function reconstructions are negligible.}.

\begin{figure}[htb]
\centerline{\includegraphics[width=0.5\textwidth]{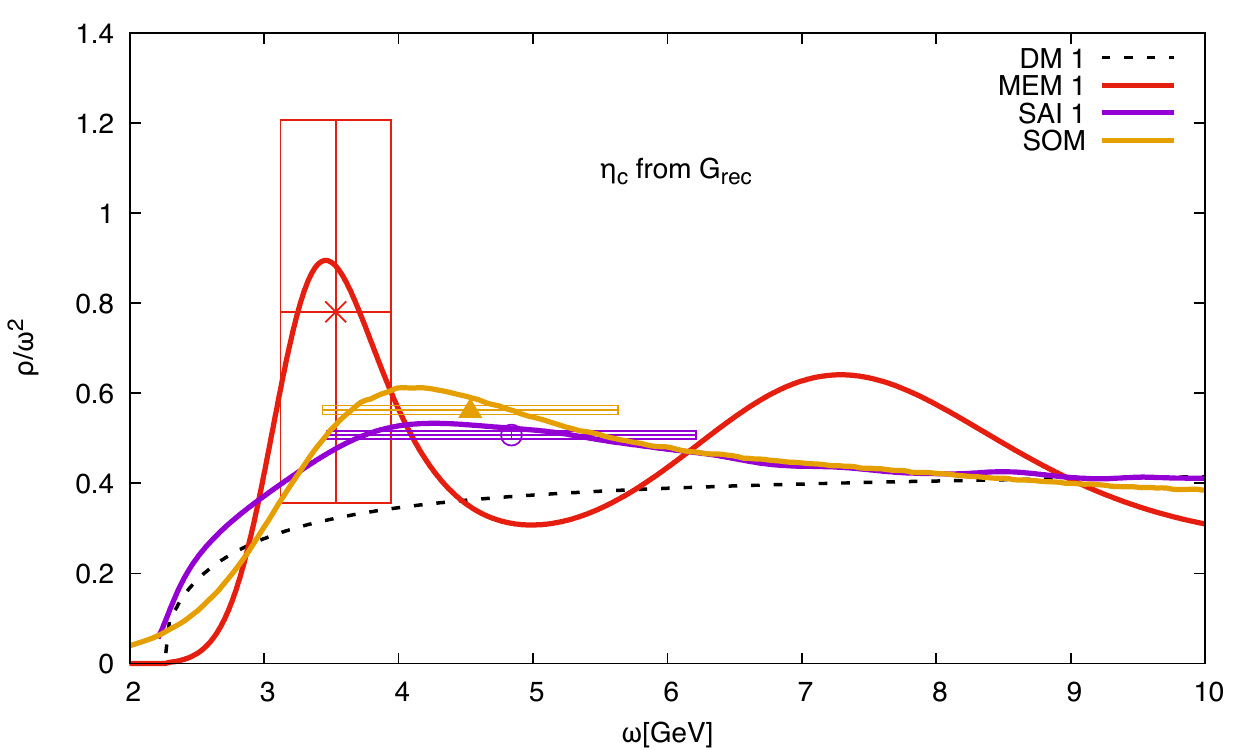} \includegraphics[width=0.5\textwidth]{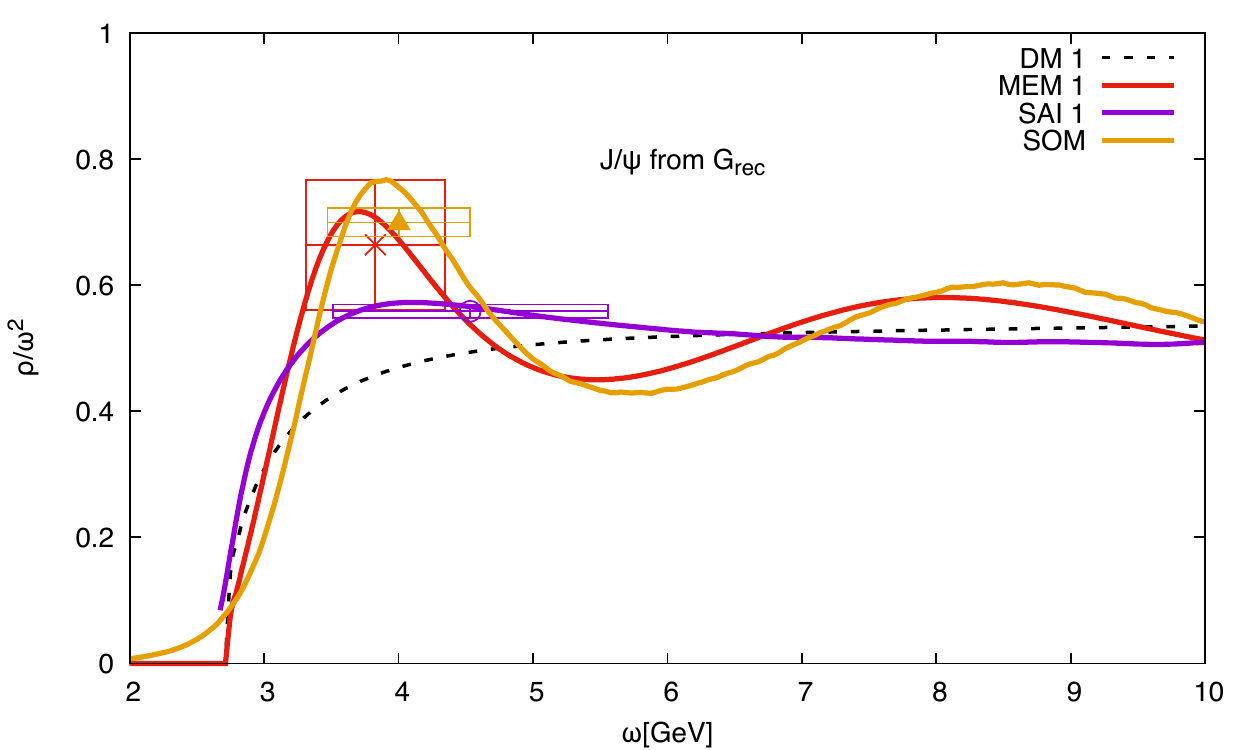}}
\caption{Significance of the strength of the resonancelike peak structure in the $\eta_c$ (\textit{left}) and $J/\psi$ (\textit{right}) spectral functions obtained using SOM, SAI and MEM from the reconstructed correlators.}
\label{error_analysis_reconstructed}
\end{figure}

Further, in order to carefully examine whether SAI, SOM and MEM indeed give different results for the reconstructed correlation functions we look into the significance of the strengths of the resonancelike structures in this case too. These results are shown in Fig.~\ref{error_analysis_reconstructed}. It seems that within one-sigma all the methods give similar results for the resonancelike peaks whose significance over the continuumlike structures  becomes questionable. Thus, we conclude that none of these methods satisfactorily reproduce the spectral function corresponding to that at $0.75T_c$ when extracted from the reconstructed correlation functions. It is not really unexpected as with similar noise-to-sginal ratios the results obtained from correlators with $N_\tau$=48 and $96$ are quite compatible with the mock results shown in the top left and bottom middle plots in Fig.~\ref{Dependence_N_noise}, respectively.

\section{Summary}
\label{conclusion}

We presented two stochastic methods, i.e. SOM and SAI, to extract spectral functions from correlation functions computed using lattice QCD. 
The SAI is a generalized stochastic method that becomes to MEM in its mean field limit. 
The other one, SOM, also a special case of SAI, does not need any default models as inputs. To test the reliability of these two methods we have tested those using various model charmonia spectral functions to mimic the cases at temperatures below and above the critical temperature. Based on these model spectral functions we computed the model correlators with different number of the data points in the temporal direction and different noise to signal ratios.
We applied SOM and SAI to these model correlators and studied in detail the dependencies of the output spectral function on the default models, number of data points as well as the noise-to-signal ratio. We found that at temperatures below the critical temperature the peak location of the first resonance peak can be correctly reproduced.
While at temperatures above the critical temperature the extraction becomes more difficult due to the additional contribution to the correlators from a transport peak in the small energy region. Extraction of the transport peak is more involved 
and it is reliable when the width of the transport peak is known. We confronted the output spectral functions obtained using SOM and SAI to those obtained from MEM, however, SOM and SAI did not show any obvious advantage over MEM in our tests. Results obtained from SOM and SAI are quite compatible with those obtained from MEM. The consistency among the results obtained from these three methods suggests that the uncertainties in the extraction of spectral functions are under control at the model data level.

In contrast to MEM, our current implementations of SAI and SOM are parallelized over many computing cores. This allows us to carry out a typical analysis within comparable wall-clock times for the methods \footnote{ However, in terms of computing cost a typical analysis for MEM, SOM and SAI take 0.4 (0.03),  320 (130) and 1200 (530)core-hours of a single Intel Xeon E5-2670 core, respectively,  for $N_\tau$ = 96 (48). The computing costs listed here are based on calculations for 420 $\alpha$ values. For the stochastic methods the computing costs also depend on the number of bases, i.e. delta functions or rectangles, as well as details of Monte Carlo samplings. We used 10000 bases and sampled 100 configurations at every 100th steps after 10000 steps for thermalization, where 1 step means the number of bases (= 10000) updates. The computing cost for SAI consists of two parts: one is generating configurations and the other is the calculation of the density of states to get $P[\alpha|G,D]$. The latter takes almost 90\% of the total computing time.}.

We also applied these methods to charmonium correlation functions in the pseudoscalar and vector channels computed on the large quenched lattice using clover-improved Wilson fermions at 0.75$T_c$ with $N_\tau=96$ and at 1.5$T_c$ with $N_\tau=48$. Even in these cases, we found consistent results using all three methods. While the location of the first resonance peak at $T<T_c$ is correctly reproduced, the location of the first bump at $1.5T_c$ is shifted to a higher frequency region by around 30\%-40\%.  However, given the fact that all three methods fail to satisfactorily reproduce the $0.75T_c$ spectral function extracted from the reconstructed correlation function, i.e. when convoluted with the integrand kernel at $1.5T_c$ having half the extent in the temporal direction, we cannot come to a definite conclusion on whether $\eta_c$ and $J/\psi$ exist as bound states in a gluon plasma at $1.5T_c$.      
With these inversion methods, in the near future, we will further refine the current results by extracting charmonium as well as bottomonium spectral functions from continuum-extrapolated correlator data with much better quality~\cite{Ding:2017rty,Burnier:2017bod}.

\section*{Acknowledgements}
\label{acknowledgements}
The work is partly supported by the National Natural Science Foundation of China under Grants No. 11775096 and No. 11535012, the Deutsche Forschungsgemeinschaft (DFG) through the Grant No. CRC-TR 211 ``Strong-interaction matter under extreme conditions’’, and by the U.S. Department of Energy, Office of Nuclear Physics through the Contract No. DE-SC001270 and Scientific Discovery through Advance Computing (ScIDAC) award "Computing the Properties of Matter with Leadership Computing Resources".  Our analyses have been done on the OCuLUS cluster at Paderborn Center for Parallel Computing and the GPU cluster at Bielefeld University. We also thank an anonymous referee whose constructive comments and suggestions helped us to significantly improve the quality and conclusion of the paper.  

\appendix
\section{A Closer Look at $P[\alpha|G,D]$}
\label{AppendixA}
Before going into detailed calculations we can obtain a qualitative conclusion on the peak position of the probability $P[\alpha|G,D]$. Assuming $P[\alpha|D]=\alpha^{-p}(p=0,1)$ the first derivative of $P[\alpha|G,D]$ with respect to $\alpha$ can be given by
\begin{equation}
\label{first_order_derivative}
\frac{\partial P[\alpha|G,D]}{\partial \alpha}\thicksim \alpha^{-N/2-p-2}Z[\langle \chi^2 \rangle _{\alpha}-(\frac{N}{2}+p)\alpha].
\end{equation}
Similarly the second derivative of $P[\alpha|G,D]$ with respect to $\alpha$ can be given by
\begin{equation}
\label{second_order_derivative}
\begin{split}
&\frac{\partial^2 P[\alpha|G,D]}{\partial^2 \alpha}\thicksim \alpha^{-N/2-p-4}Z_{\alpha}\times\\
&[(\frac{N}{2}+p)(\frac{N}{2}+p+1)\alpha^2-2\alpha(\frac{N}{2}+p+1)\langle \chi^2 \rangle _{\alpha}-\langle (\chi^2)^2 \rangle _{\alpha}].
\end{split}
\end{equation}
Since $Z_{\alpha}$ is always positive at any $\alpha > 0$ by definition, Eq.(\ref{first_order_derivative}) becomes zero if or only if
\begin{equation}
\label{first_order_derivative_0}
\langle \chi^2 \rangle _{\alpha}-\alpha(\frac{N}{2}+p)=0 \leftrightarrow \langle \frac{\chi^2}{N} \rangle _{\alpha}-\alpha(\frac{p}{N}+\frac{1}{2})=0,
\end{equation}
which means that $P[\alpha|G,D]$ has an extrema at $\alpha^*$ where the condition Eq.(\ref{first_order_derivative_0}) is satisfied. And we also learn that for sufficiently large $N$ , the difference in cases for $p = 0$ and 1 is negligible. At $\alpha=\alpha^*$ where
\begin{equation}
\label{second_order_derivative_0}
\begin{split}
&\frac{\partial^2 P[\alpha|G,D]}{\partial^2 \alpha}\Big{|}_{\alpha=\alpha^*}\thicksim -(\alpha^*)^{-N/2-p-4}Z_{\alpha^*}[\langle (\chi^2)^2 \rangle _{\alpha}+\langle \chi^2 \rangle ^2_{\alpha}+\alpha^*\langle \chi^2\rangle_{\alpha^*}],
\end{split}
\end{equation}
$\langle (\chi^2)^2 \rangle _{\alpha}+\langle \chi^2 \rangle ^2_{\alpha}+\alpha^*\langle \chi^2\rangle_{\alpha^*}>0$ is always satisfied which means the extreme value at $\alpha^*$ is a maximum.

\section{Wang-Landau Algorithm}
\label{AppendixB}
The Wang-Landau algorithm is a Monte Carlo method to compute density of states $\Omega(E)$ of a system. An ordinary Metropolis algorithm, which samples with Boltzmann weights $e^{-E/\alpha}$, can only generate the distribution of $\Omega(E)e^{-E/\alpha}$ at a fixed temperature $\alpha$, while WLA calculates $\Omega(E)$ directly in the whole energy range and hence, $\Omega(E)e^{-E/\alpha}$ at any temperature can be constructed accordingly. In WLA what one needs to do is to produce a flat histogram by a random walk in the energy space with the probability proportional to $1/\Omega(E)$ for visiting an energy level $E$. Given the energy of the system before the walk as $E_i$ and after that $E_j$, after a tremendous number of iterations we would arrive at a ``flat histogram" if we accept the walk with a transition probability $p_{i\rightarrow j}=\min \big [1, \frac{\Omega(E_i)}{\Omega({E_j})}]$.

At the beginning of a simulation we do not know the density of states $\Omega(E)$ $a\ priori$, we just let all $\Omega(E)$ equal to a same constant, in our case 1. And set the histogram count $h(E)=0$ for all energy levels. Once the walk to $E_j$ is accepted we increase the histogram count $h(E_j)$ by 1 and update the DoS by
\begin{align}
\label{update_dos}
\Omega(E_j) \rightarrow \Omega(E_j) \times f,
\end{align}
where $f$ is a controlling factor. Otherwise we update the previous energy level $E_i$ in the same way. For each $f$, we perform successive walks until the flatness criterion is satisfied. That is, $\frac{h(E)}{\langle h(E)\rangle}\geq x$ is satisfied for all $E$ where ${\langle h(E)\rangle}$ is the mean value of $h(E)$ and $0<x<1$ is a flatness parameter. Once the flatness criterion is satisfied we modify the factor $f$ by $f \rightarrow f^p $ where $0<p<1$ and reset the histogram counts to zero. In the next iteration we repeat the previous procedure and finally when $f$ reaches the predefined value $f_{stop}$, the DoS converges very close to its true value. The accuracy of DoS obtained from WLA depends on the parameters in the simulation like $x$, $p$, $f_{stop}$, etc. With larger $x$ and $p$ or smaller $f_{stop}$, the DoS obtained is more accurate but apparently this makes the convergence slower.

For the system interested, i.e. a configuration consisting of many delta functions, the energy range is quite wide, usually from 1 to $10^7$ or larger. Besides, the information from a smaller energy range is more important. In such a large range we cannot divide the energy $E$ uniformly. We thus generalize the flat criterion to $\frac{\Omega(E_i)}{\Omega'(E_i) }\Delta E_i\sim h(E_i)=const$. Here $\Omega(E_i)$ is the true DoS while $\Omega'(E_i)$ is the estimated DoS. If we divide $E$ uniformly, $\Delta E_i$ is the same for every $E_i$. In our case, we divide $\ln (E)$ uniformly and denote $\ln \Omega(E)$ as $g(E)$, then we can get $g(E_i)=const+g'(E_i)-\delta\cdot i$ readily if we denote $\delta=\frac{\ln E_{max}-\ln E_{min}}{N_E}$ where $N_E$ is the number of energy levels.

The replica exchange in Wang–Landau sampling is a parallel version of WLA that provides us the opportunity to implement WLA on massively parallel supercomputers. In this scheme the energy range is divided into overlapped subwindows. In each subwindow the standard WLA is performed. The difference is after a certain number of walks, a replica exchange step is carried out between the neighbouring subwindows. Denote $E(x)$ and $E(Y)$ as the energy of the configurations to be exchanged in two neighbouring subwindows, say $a$ and $b$, and the corresponding logarithm of the density of states before exchange are $g_a(E(X))$ and $g_b(E(Y))$, respectively. Suppose after the exchange the logarithm of the density of states become $g_a(E(Y))$ and $g_b(E(X))$. The probability to accept the exchange of configurations is
\begin{align}
\label{probabililty_replica}
P_{acc}=\min \Big{[}1,\frac{g_a(E(X))}{g_a(E(Y))}\frac{g_b(E(Y))}{g_b(E(X))} \Big{]}.
\end{align}
If after the replica exchange either $E(x)$ goes outside the range of subwindow $b$ or $E(Y)$ goes outside the range of subwindow $a$, the replica exchange fails. Then we just discard the exchange and move on to the next step. After the predefined $f_{stop}$ is reached, we join the DoS in each subwindows to obtain the DoS in the whole energy range.

\begin{figure}[htb]
\centerline{\includegraphics[width=0.45\textwidth]{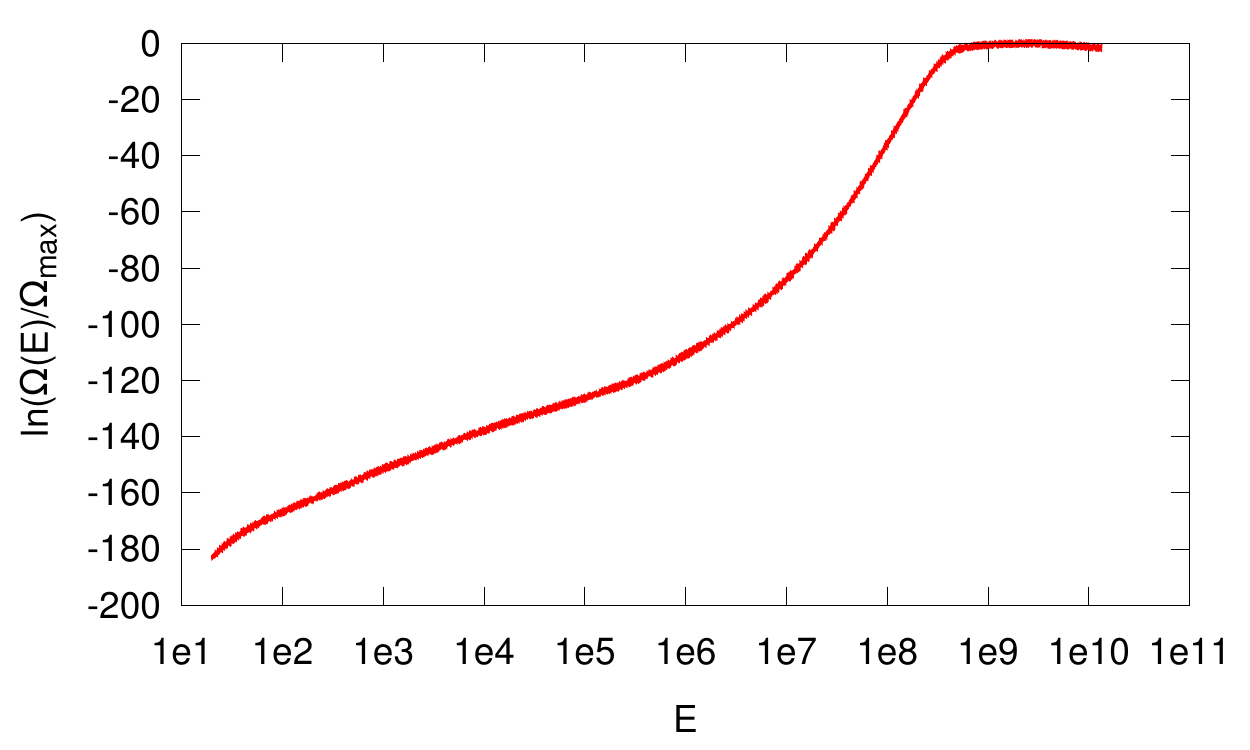} \includegraphics[width=0.45\textwidth]{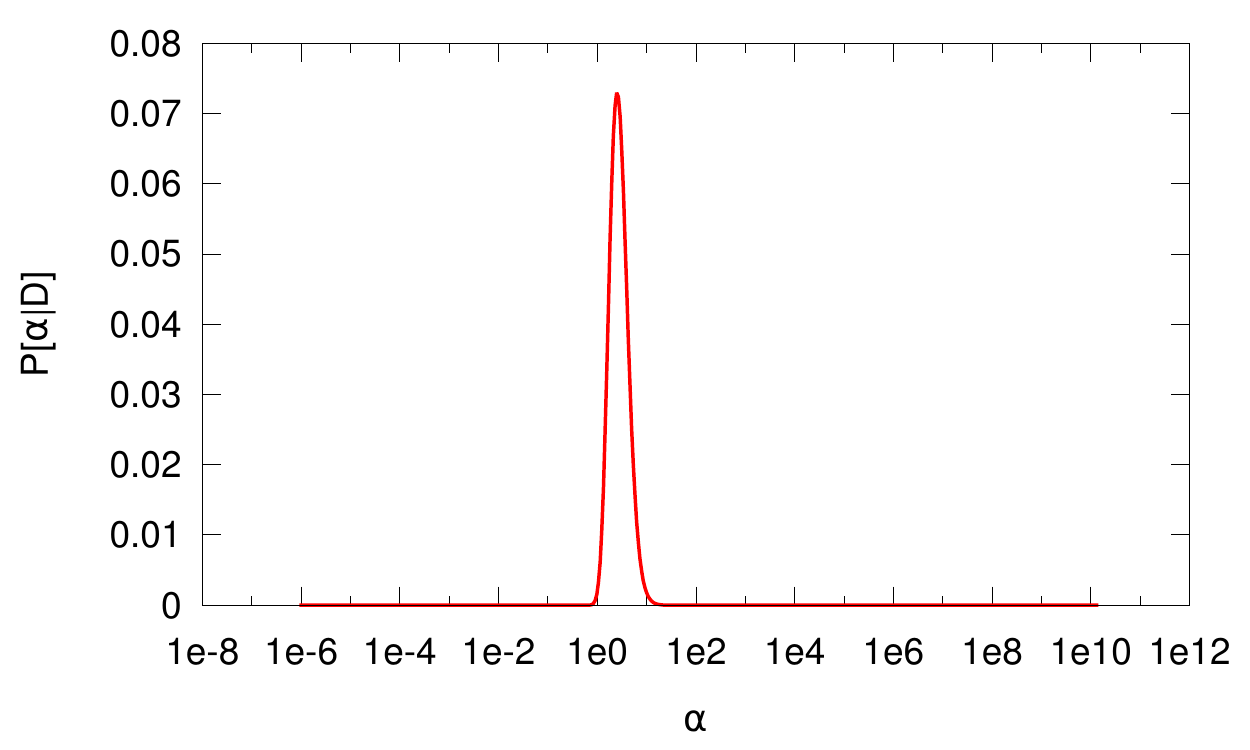}}
\caption{\textit{Left}: DoS calculated by replica exchange Wang-Landau sampling. \textit{Right}: A typical distribution of $P[\alpha|G,D]$. Both are from model data tests.}
\label{DoS_pag1}
\end{figure}

Finally as an illustration we show the DoS calculated using model data on the left in Fig.\ref{DoS_pag1} and $P[\alpha|G,D]$ calculated accordingly is shown on the right. The flatness parameter in WLA is set to 0.85 and $\ln f_{stop}$ to $10^{-7}$. The entire energy domain is divided into 600 subdomains with $25\%$ overlap on both sides, where each subdomain has 24 energy levels. The parameters used in the model spectral function can be found in Table~\ref{Mock_SPFs_Parameters}.

\bibliographystyle{JHEP}
\bibliography{bibfile}

\providecommand{\href}[2]{#2}\begingroup\raggedright\begin{thebibliography}{10}

\bibitem{Aoki:2006we}
Y.~Aoki, G.~Endrodi, Z.~Fodor, S.~D. Katz and K.~K. Szabo, \emph{{The Order of
  the quantum chromodynamics transition predicted by the standard model of
  particle physics}}, \href{https://doi.org/10.1038/nature05120}{\emph{Nature}
  {\bfseries 443} (2006) 675--678},
  [\href{https://arxiv.org/abs/hep-lat/0611014}{{\ttfamily hep-lat/0611014}}].

\bibitem{Bazavov:2011nk}
A.~Bazavov et~al., \emph{{The chiral and deconfinement aspects of the QCD
  transition}}, \href{https://doi.org/10.1103/PhysRevD.85.054503}{\emph{Phys.
  Rev.} {\bfseries D85} (2012) 054503},
  [\href{https://arxiv.org/abs/1111.1710}{{\ttfamily 1111.1710}}].

\bibitem{Brandenburg:2017meb}
{\scshape STAR} collaboration, J.~D. Brandenburg, \emph{{Dilepton Production in
  p$+$p, Au$+$Au collisions at $\sqrt{s_{NN}}$ = 200 GeV and U$+$U collisions
  at $\sqrt{s_{NN}}$ = 193 GeV}},
  \href{https://doi.org/10.1016/j.nuclphysa.2017.05.093}{\emph{Nucl. Phys.}
  {\bfseries A967} (2017) 676--679},
  [\href{https://arxiv.org/abs/1704.06890}{{\ttfamily 1704.06890}}].

\bibitem{Adamczyk:2015lme}
{\scshape STAR} collaboration, L.~Adamczyk et~al., \emph{{Measurements of
  Dielectron Production in Au$+$Au Collisions at $\sqrt{s_{\rm NN}}$ = 200 GeV
  from the STAR Experiment}},
  \href{https://doi.org/10.1103/PhysRevC.92.024912}{\emph{Phys. Rev.}
  {\bfseries C92} (2015) 024912},
  [\href{https://arxiv.org/abs/1504.01317}{{\ttfamily 1504.01317}}].

\bibitem{Adare:2015ila}
{\scshape PHENIX} collaboration, A.~Adare et~al., \emph{{Dielectron production
  in Au$+$Au collisions at $\sqrt{s_{NN}}$=200 GeV}},
  \href{https://doi.org/10.1103/PhysRevC.93.014904}{\emph{Phys. Rev.}
  {\bfseries C93} (2016) 014904},
  [\href{https://arxiv.org/abs/1509.04667}{{\ttfamily 1509.04667}}].

\bibitem{ALICE:2012ab}
{\scshape ALICE} collaboration, B.~Abelev et~al., \emph{{Suppression of high
  transverse momentum D mesons in central Pb-Pb collisions at
  $\sqrt{s_{NN}}=2.76$ TeV}},
  \href{https://doi.org/10.1007/JHEP09(2012)112}{\emph{JHEP} {\bfseries 09}
  (2012) 112}, [\href{https://arxiv.org/abs/1203.2160}{{\ttfamily 1203.2160}}].

\bibitem{Adare:2006ns}
{\scshape PHENIX} collaboration, A.~Adare et~al., \emph{{$J/\psi$ Production vs
  Centrality, Transverse Momentum, and Rapidity in Au+Au Collisions at
  $\sqrt{s_{NN}} = 200$ GeV}},
  \href{https://doi.org/10.1103/PhysRevLett.98.232301}{\emph{Phys. Rev. Lett.}
  {\bfseries 98} (2007) 232301},
  [\href{https://arxiv.org/abs/nucl-ex/0611020}{{\ttfamily nucl-ex/0611020}}].

\bibitem{Chatrchyan:2012np}
{\scshape CMS} collaboration, S.~Chatrchyan et~al., \emph{{Suppression of
  non-prompt $J/\psi$, prompt $J/\psi$, and Y(1S) in PbPb collisions at
  $\sqrt{s_{NN}}=2.76$ TeV}},
  \href{https://doi.org/10.1007/JHEP05(2012)063}{\emph{JHEP} {\bfseries 05}
  (2012) 063}, [\href{https://arxiv.org/abs/1201.5069}{{\ttfamily 1201.5069}}].

\bibitem{Braaten:1990wp}
E.~Braaten, R.~D. Pisarski and T.-C. Yuan, \emph{{Production of Soft Dileptons
  in the Quark - Gluon Plasma}},
  \href{https://doi.org/10.1103/PhysRevLett.64.2242}{\emph{Phys. Rev. Lett.}
  {\bfseries 64} (1990) 2242}.

\bibitem{Ding:2015ona}
H.-T. Ding, F.~Karsch and S.~Mukherjee, \emph{{Thermodynamics of
  strong-interaction matter from Lattice QCD}},
  \href{https://doi.org/10.1142/S0218301315300076}{\emph{Int. J. Mod. Phys.}
  {\bfseries E24} (2015) 1530007},
  [\href{https://arxiv.org/abs/1504.05274}{{\ttfamily 1504.05274}}].

\bibitem{Ghiglieri:2016tvj}
J.~Ghiglieri, O.~Kaczmarek, M.~Laine and F.~Meyer, \emph{{Lattice constraints
  on the thermal photon rate}},
  \href{https://doi.org/10.1103/PhysRevD.94.016005}{\emph{Phys. Rev.}
  {\bfseries D94} (2016) 016005},
  [\href{https://arxiv.org/abs/1604.07544}{{\ttfamily 1604.07544}}].

\bibitem{Ding:2016hua}
H.-T. Ding, O.~Kaczmarek and F.~Meyer, \emph{{Thermal dilepton rates and
  electrical conductivity of the QGP from the lattice}},
  \href{https://doi.org/10.1103/PhysRevD.94.034504}{\emph{Phys. Rev.}
  {\bfseries D94} (2016) 034504},
  [\href{https://arxiv.org/abs/1604.06712}{{\ttfamily 1604.06712}}].

\bibitem{Asakawa:2000tr}
M.~Asakawa, T.~Hatsuda and Y.~Nakahara, \emph{{Maximum entropy analysis of the
  spectral functions in lattice QCD}},
  \href{https://doi.org/10.1016/S0146-6410(01)00150-8}{\emph{Prog.Part.Nucl.Phys.}
  {\bfseries 46} (2001) 459--508},
  [\href{https://arxiv.org/abs/hep-lat/0011040}{{\ttfamily hep-lat/0011040}}].

\bibitem{Aarts:2007pk}
G.~Aarts, C.~Allton, M.~B. Oktay, M.~Peardon and J.-I. Skullerud,
  \emph{{Charmonium at high temperature in two-flavor QCD}},
  \href{https://doi.org/10.1103/PhysRevD.76.094513}{\emph{Phys. Rev.}
  {\bfseries D76} (2007) 094513},
  [\href{https://arxiv.org/abs/0705.2198}{{\ttfamily 0705.2198}}].

\bibitem{Ikeda:2016czj}
A.~Ikeda, M.~Asakawa and M.~Kitazawa, \emph{{In-medium dispersion relations of
  charmonia studied by maximum entropy method}},
  \href{https://doi.org/10.1103/PhysRevD.95.014504}{\emph{Phys. Rev.}
  {\bfseries D95} (2017) 014504},
  [\href{https://arxiv.org/abs/1610.07787}{{\ttfamily 1610.07787}}].

\bibitem{Burnier:2013nla}
Y.~Burnier and A.~Rothkopf, \emph{{Bayesian Approach to Spectral Function
  Reconstruction for Euclidean Quantum Field Theories}},
  \href{https://doi.org/10.1103/PhysRevLett.111.182003}{\emph{Phys. Rev. Lett.}
  {\bfseries 111} (2013) 182003},
  [\href{https://arxiv.org/abs/1307.6106}{{\ttfamily 1307.6106}}].

\bibitem{Francis:2015daa}
A.~Francis, O.~Kaczmarek, M.~Laine, T.~Neuhaus and H.~Ohno,
  \emph{{Nonperturbative estimate of the heavy quark momentum diffusion
  coefficient}}, \href{https://doi.org/10.1103/PhysRevD.92.116003}{\emph{Phys.
  Rev.} {\bfseries D92} (2015) 116003},
  [\href{https://arxiv.org/abs/1508.04543}{{\ttfamily 1508.04543}}].

\bibitem{Brandt:2015aqk}
B.~B. Brandt, A.~Francis, B.~Jäger and H.~B. Meyer, \emph{{Charge transport
  and vector meson dissociation across the thermal phase transition in lattice
  QCD with two light quark flavors}},
  \href{https://doi.org/10.1103/PhysRevD.93.054510}{\emph{Phys. Rev.}
  {\bfseries D93} (2016) 054510},
  [\href{https://arxiv.org/abs/1512.07249}{{\ttfamily 1512.07249}}].

\bibitem{Dudal:2013yva}
D.~Dudal, O.~Oliveira and P.~J. Silva, \emph{{Källén-Lehmann spectroscopy for
  (un)physical degrees of freedom}},
  \href{https://doi.org/10.1103/PhysRevD.89.014010}{\emph{Phys. Rev.}
  {\bfseries D89} (2014) 014010},
  [\href{https://arxiv.org/abs/1310.4069}{{\ttfamily 1310.4069}}].

\bibitem{PhysRevE.81.056701}
S.~Fuchs, T.~Pruschke and M.~Jarrell, \emph{Analytic continuation of quantum
  monte carlo data by stochastic analytical inference},
  \href{https://doi.org/10.1103/PhysRevE.81.056701}{\emph{Phys. Rev. E}
  {\bfseries 81} (May, 2010) 056701}.

\bibitem{2004cond.mat..3055B}
K.~S.~D. {Beach}, \emph{{Identifying the maximum entropy method as a special
  limit of stochastic analytic continuation}}, {\emph{eprint
  arXiv:cond-mat/0403055} (Mar., 2004) }.

\bibitem{PhysRevB.62.6317}
A.~S. Mishchenko, N.~V. Prokof'ev, A.~Sakamoto and B.~V. Svistunov,
  \emph{Diagrammatic quantum monte carlo study of the fr\"ohlich polaron},
  \href{https://doi.org/10.1103/PhysRevB.62.6317}{\emph{Phys. Rev. B}
  {\bfseries 62} (Sep, 2000) 6317--6336}.

\bibitem{Ding:2012sp}
H.~T. Ding, A.~Francis, O.~Kaczmarek, F.~Karsch, H.~Satz and W.~Soeldner,
  \emph{{Charmonium properties in hot quenched lattice QCD}},
  \href{https://doi.org/10.1103/PhysRevD.86.014509}{\emph{Phys. Rev.}
  {\bfseries D86} (2012) 014509},
  [\href{https://arxiv.org/abs/1204.4945}{{\ttfamily 1204.4945}}].

\bibitem{Jarrell:1996rrw}
M.~Jarrell and J.~E. Gubernatis, \emph{{Bayesian inference and the analytic
  continuation of imaginary-time quantum Monte Carlo data}},
  \href{https://doi.org/10.1016/0370-1573(95)00074-7}{\emph{Phys. Rept.}
  {\bfseries 269} (1996) 133--195}.

\bibitem{Jeffreys:1998}
H.~Jeffreys{\emph{Theory of Probability(Third Edition),(Oxford Univ. Press,
  Oxford, 1998)} }.

\bibitem{Box:1992}
G.~E.~P. Box and G.~C. Tiao{\emph{Bayesian Inference in Statistical
  Analysis,(John Wiley and Sons, New York, 1992)} }.

\bibitem{PhysRevE.90.023302}
T.~Vogel, Y.~W. Li, T.~W\"ust and D.~P. Landau, \emph{Scalable replica-exchange
  framework for wang-landau sampling},
  \href{https://doi.org/10.1103/PhysRevE.90.023302}{\emph{Phys. Rev. E}
  {\bfseries 90} (Aug, 2014) 023302}.

\bibitem{Marinari:1996dh}
E.~Marinari, \emph{{Optimized Monte Carlo methods}},
  \href{https://arxiv.org/abs/cond-mat/9612010}{{\ttfamily cond-mat/9612010}}.

\bibitem{Karsch:2003wy}
F.~Karsch, E.~Laermann, P.~Petreczky and S.~Stickan, \emph{{Infinite
  temperature limit of meson spectral functions calculated on the lattice}},
  \href{https://doi.org/10.1103/PhysRevD.68.014504}{\emph{Phys. Rev.}
  {\bfseries D68} (2003) 014504},
  [\href{https://arxiv.org/abs/hep-lat/0303017}{{\ttfamily hep-lat/0303017}}].

\bibitem{Aarts:2005hg}
G.~Aarts and J.~M. Martinez~Resco, \emph{{Continuum and lattice meson spectral
  functions at nonzero momentum and high temperature}},
  \href{https://doi.org/10.1016/j.nuclphysb.2005.08.012}{\emph{Nucl. Phys.}
  {\bfseries B726} (2005) 93--108},
  [\href{https://arxiv.org/abs/hep-lat/0507004}{{\ttfamily hep-lat/0507004}}].

\bibitem{Umeda:2007hy}
T.~Umeda, \emph{{A Constant contribution in meson correlators at finite
  temperature}}, \href{https://doi.org/10.1103/PhysRevD.75.094502}{\emph{Phys.
  Rev.} {\bfseries D75} (2007) 094502},
  [\href{https://arxiv.org/abs/hep-lat/0701005}{{\ttfamily hep-lat/0701005}}].

\bibitem{Ding:2017rty}
H.-T. Ding, O.~Kaczmarek, A.-L. Kruse, H.~Ohno and H.~Sandmeyer,
  \emph{{Continuum extrapolation of quarkonium correlators at non-zero
  temperature}},  \href{https://arxiv.org/abs/1710.08858}{{\ttfamily
  1710.08858}}.

\bibitem{Burnier:2017bod}
Y.~Burnier, H.~T. Ding, O.~Kaczmarek, A.~L. Kruse, M.~Laine, H.~Ohno et~al.,
  \emph{{Thermal quarkonium physics in the pseudoscalar channel}},
  \href{https://arxiv.org/abs/1709.07612}{{\ttfamily 1709.07612}}.

\end{thebibliography}\endgroup

\end{document}